\documentclass[sigplan,10pt,nonacm]{acmart}

\usepackage[utf8]{inputenc} 
\usepackage[T1]{fontenc} 
\usepackage{microtype} 
\usepackage{fancyhdr}  
\usepackage{color}
\usepackage{xcolor}
\usepackage{etoolbox}

\usepackage{algorithm} 

\usepackage{mathtools}
\usepackage{mathptmx}

\usepackage{amsmath}
\usepackage{amsfonts}

\usepackage{bm}
\usepackage{enumerate}
\usepackage{lscape}
\usepackage{rotating}
\usepackage{xspace}
\usepackage{relsize}

\usepackage{multirow}
\usepackage{multicol}
\usepackage{booktabs} 
\usepackage{longtable}
\usepackage{array}
\usepackage{colortbl}

\usepackage{url}
\usepackage{textcomp}

\usepackage{wrapfig}
\usepackage[framemethod=TikZ]{mdframed}
\usepackage{tikz}
\usetikzlibrary{arrows,shapes,decorations,automata,backgrounds,calc,topaths,positioning,matrix}
\usepackage{lipsum}
\usepackage{balance}
\usepackage{listings}
\usepackage{graphicx}
\usepackage[normalem]{ulem}

\usepackage{subcaption}
\usepackage{hyperref}
\hypersetup{urlcolor=black,colorlinks=false,hidelinks,linkcolor=blue,filecolor=magenta,urlcolor=cyan}
\usepackage{cleveref}
\usepackage{relsize}

\usepackage[most]{tcolorbox}
\usepackage{listings}
\usepackage{textcomp}
\usepackage{xspace}

\newcommand{\originalgrumbler}[2]{\begin{quote}\textcolor{blue}{\sl{\bf #1 says:} #2}\end{quote}}
\newcommand{\grumbler}[2]{\originalgrumbler{#1}{#2}}
\newcommand{\swarnendu}[1]{\grumbler{Swarnendu}{#1}}

\newcommand{\sagnik}[1]{\grumbler{Sagnik}{#1}}
\newcommand{\mayant}[1]{\grumbler{Mayant}{#1}}

\newcommand{\later}[1]{\begin{quote}\textcolor{darkgreen}{\textbackslash \textbf{later\{}} #1 \textcolor{darkgreen}{\}}\end{quote}}
\newcommand{\revise}[1]{\textcolor{orange}{REVISE: #1}}

\definecolor{darkgreen}{rgb}{0,0.4,0}
\definecolor{mygreen}{rgb}{0,0.6,0}
\definecolor{mygray}{rgb}{0.5,0.5,0.5}
\definecolor{mymauve}{rgb}{0.58,0,0.82}
\definecolor{backgroundColour}{rgb}{0.95,0.95,0.92}

\lstdefinestyle{GenericStyle}{
    frame=tb, 
    backgroundcolor=\color{backgroundColour},
    basicstyle=\small,
    keywordstyle=\color{black}\textbf,
    stringstyle=\ttfamily,
    commentstyle=\color{darkgreen},
    showlines=true, 
    numbers=left,
    numberstyle=\tiny,
    stepnumber=1,
    numbersep=5pt,
    tabsize=2,
    showstringspaces=false
}

\lstdefinestyle{BashStyle}{
    language=bash,
    frame=tb, 
    backgroundcolor=\color{backgroundColour},
    basicstyle=\small\sffamily,
    linewidth=1.0\linewidth,
    xleftmargin=0.1\linewidth,
    showlines=true, 
    numbers=left,
    numberstyle=\tiny,
    stepnumber=1,
    numbersep=5pt,
    columns=fullflexible
}

\lstdefinestyle{JavaStyle}{
    language=Java, 
    frame=tb,
    backgroundcolor=\color{backgroundColour},
    basicstyle=\footnotesize\sffamily,
    keywordstyle=\color{blue}\textbf,
    commentstyle=\color{darkgreen}\textit,
    numbers=left,
    numberstyle=\tiny,
    stepnumber=1,
    numbersep=5pt,
    tabsize=2,
    extendedchars=true,
    breaklines=true,
    escapeinside={\%*}{*)}, 
    columns=flexible
}

\lstdefinestyle{CPPStyle}{
    language=C++,
    frame=tb, 
    backgroundcolor=\color{backgroundColour},
    basicstyle=\footnotesize\ttfamily,
    keywordstyle=\color{blue}\textbf,
    stringstyle=\color{red}\ttfamily,
    commentstyle=\color{darkgreen}\textit,
    morecomment=[l][\color{magenta}]{\#},
    literate = *{\ \ }{\ }1, 
    showlines=true, 
    numbers=left,
    numberstyle=\tiny,
    stepnumber=1,
    numbersep=5pt,
    tabsize=2
}

\lstdefinestyle{CStyle}{
    language=C,
    frame=tb, 
    backgroundcolor=\color{backgroundColour},
    basicstyle=\footnotesize\ttfamily,
    stringstyle=\color{mymauve},
    commentstyle=\color{mygreen}\textit,
    keywordstyle=\color{magenta},
    numberstyle=\scriptsize\color{mygray},
    breaklines=true,
    showlines=true, 
    numbers=left,
    numbersep=5pt,
    stepnumber=1,
    tabsize=2,
    showstringspaces=false
}

\lstdefinestyle{PythonStyle}{
    language=Python,
    frame=tb, 
    backgroundcolor=\color{backgroundColour},
    basicstyle=\ttm,
    keywordstyle=\ttb\color{deepblue},
    stringstyle=\color{deepgreen},
    otherkeywords={self},             
    emph={MyClass,__init__},          
    emphstyle=\ttb\color{deepred},    
    showstringspaces=false
}

\lstdefinestyle{CUDAStyle}{
    language=C++,
    frame=tb, 
    backgroundcolor=\color{backgroundColour},
    basicstyle=\footnotesize\ttfamily,
    keywordstyle=\color{blue}\textbf,
    stringstyle=\color{red}\ttfamily,
    commentstyle=\color{darkgreen}\textit,
    emph={
            cudaMalloc, cudaFree, __global__, __shared__, __device__, __host__, __syncthreads,
        },
    emphstyle=\color{darkgreen}\bf\ttfamily,
    numberstyle=\tiny\color{mygray},
    breaklines=true,
    showlines=true, 
    moredelim=[s][\ttfamily]{<<<}{>>>},
    numbers=left,
    numbersep=3pt,
    stepnumber=1,
    tabsize=2,
    showstringspaces=false
}

\newcommand{\bench}[1]{\textsf{\small #1}}
\newcommand{\code}[1]{\texttt{\small #1}}
\newcommand{\pcode}[1]{\textsf{#1}}

\newcommand{\eg}{e.g.\xspace}
\newcommand{\ie}{i.e.\xspace}

\newcommand{\naive}{na\"{\i}ve\xspace}

\newcommand{\naively}{na\"{\i}vely\xspace}

\tcbset{
    frame code={}
    center title,
    left=0pt,
    right=0pt,
    top=0pt,
    bottom=0pt,
    colback=gray!20,
    colframe=white,
    width=\dimexpr\textwidth\relax,
    enlarge left by=0mm,
    boxsep=5pt,
    arc=0pt,outer arc=0pt,
}

\graphicspath{{./figs/}}

\newcommand{\nvidia}{NVIDIA\xspace}
\newcommand{\NVIDIA}{NVIDIA\xspace}

\newcommand{\bigO}[1]{\( \mathcal{O} \left( #1 \right) \)}

\usepackage{amsthm}
\newtheorem{definition}{Definition}

\newcommand{\barracuda}{Barracuda\xspace}
\newcommand{\Barracuda}{\barracuda}

\newcommand{\barracudap}{Barracuda+\xspace}
\newcommand{\Barracudap}{\barracudap}

\newcommand{\scord}{ScoRD\xspace}
\newcommand{\Scord}{\scord}

\newcommand{\scor}{ScoR\xspace}

\newcommand{\approach}{PreDataR\xspace}
\newcommand{\Approach}{\approach}

\newcommand{\iguard}{iGUARD\xspace}
\newcommand{\IGUARD}{\iguard}

\lstdefinestyle{TraceStyle}{
    language=C,
    frame=t,
    backgroundcolor=\color{white},
    commentstyle=\color{mygreen}\textit,
    basicstyle=\footnotesize\ttfamily,
    keywordstyle=\color{magenta},
    numberstyle=\scriptsize\color{mygray},
    stringstyle=\color{mymauve},
    breaklines=true,
    showstringspaces=false,
}

\newcommand{\gwcp}{GWCP\xspace}
\newcommand{\GWCP}{\gwcp}

\newcommand{\conv}{\bench{1dconv}\xspace}

\newcommand{\mm}{\bench{matmul}\xspace}

\newcommand{\red}{\bench{reduction}\xspace}

\newcommand{\ruleone}{\bench{rule-110}\xspace}

\newcommand{\pf}{\bench{pathfinder}\xspace}

\newcommand{\strmcls}{\bench{streamcluster}\xspace}

\newcommand{\tfred}{\bench{threadFenceReduction}\xspace}

\newcommand{\needle}{\bench{needle}\xspace}

\newcommand{\hotspot}{\bench{hotspot}\xspace}

\newcommand{\kmeans}{\bench{kmeans}\xspace}

\newcommand{\stencil}{stencil\xspace}

\newcommand{\thr}[1]{\textsf{#1}}
\renewcommand{\pcode}[1]{\textsf{\small #1}}

\newcommand{\bcpunmodified}{72.3X\xspace}
\newcommand{\bcpblank}{5.2X\xspace}

\newcommand{\scordunmodified}{35.8X\xspace}
\newcommand{\scordblank}{2.6X\xspace}

\newcommand{\gwcpunmodified}{97.5X\xspace}
\newcommand{\gwcpblank}{7.1X\xspace}

\newtoggle{techreport}
\togglefalse{techreport}

\newtoggle{acmFormat}
\toggletrue{acmFormat}

\newtoggle{ieeeFormat}
\togglefalse{ieeeFormat}

\newtoggle{asplos}
\togglefalse{asplos}

\renewcommand{\grumbler}[2]{}
\renewcommand{\later}[1]{}

\iftoggle{ieeeFormat}{
    \usepackage{cite}
    \usepackage{amssymb}

    \usepackage{algorithm}
    \usepackage{algorithmic}

    \IEEEoverridecommandlockouts

    \def\BibTeX{{\rm B\kern-.05em{\sc i\kern-.025em b}\kern-.08em
    T\kern-.1667em\lower.7ex\hbox{E}\kern-.125emX}}

}{}

\iftoggle{acmFormat}{
    \usepackage{algorithmicx}
    \usepackage[noend]{algpseudocode}
    \algnewcommand\And{\textbf{and}\xspace}
    \algnewcommand\Or{\textbf{or}\xspace}

    \settopmatter{printacmref=false, printccs=false, printfolios=true}

    \acmConference[PLDI'22]{ACM SIGPLAN Conference on Programming Language Design and Implementation}{June 20--24, 2022}{San Diego, CA, USA}
    \acmYear{2022}
    \acmISBN{}
    \acmDOI{}
    \startPage{1}

    \setcopyright{none}

    \renewcommand\footnotetextcopyrightpermission[1]{}

}{}

\iftoggle{asplos}{
    
}{}

\begin{document}

\title{Predictive Data Race Detection for GPUs}


\iftoggle{acmFormat}{
    \author{Sagnik Dey}
    \email{mail.sagnik.dey@gmail.com}
    \affiliation{%
        \department{Department of Mathematics and Statistics}
        \institution{Indian Institute of Technology Kanpur \country{India}}}
    \authornote{Both authors contributed equally to this work.}
    \author{Mayant Mukul}
    \authornotemark[1]
    \authornote{The author contributed to the work when he was affiliated with Indian Insitute of Technology Kanpur.}
    \email{mayantmukul@gmail.com}
    \affiliation{%
        \institution{Dream11 \country{India}}}
    \author{Parth Sharma}
    \email{prthshrma20@iitk.ac.in}
    \affiliation{%
        \department{Department of Computer Science and Engineering}
        \institution{Indian Institute of Technology Kanpur \country{India}}}
    \author{Swarnendu Biswas}
    \email{swarnendu@cse.iitk.ac.in}
    \affiliation{%
        \department{Department of Computer Science and Engineering}
        \institution{Indian Institute of Technology Kanpur \country{India}}}

    \renewcommand{\shortauthors}{Sagnik Dey et al.}

    \begin{abstract}

    The high degree of parallelism and relatively complicated synchronization mechanisms in GPUs make writing correct kernels difficult. Data races pose one such concurrency correctness challenge, and therefore, effective methods of detecting as many data races as possible are required.

    Predictive partial order relations for CPU programs aim to expose data races that can be hidden
    during a dynamic execution. Existing predictive partial orders
    cannot be \naively applied to analyze GPU kernels because of the differences in programming
    models. This work proposes \emph{\gwcp}, a predictive partial order for data race detection of
    GPU kernels. \GWCP extends a sound and precise relation called weak-causally-precedes (WCP)
    proposed in the context of multithreaded shared memory CPU programs to GPU kernels. \GWCP takes into account the GPU thread hierarchy and different synchronization semantics such as barrier synchronization and scoped atomics and locks.

    We implement a tool called \approach that tracks the \gwcp relation using binary instrumentation.
    \Approach includes three optimizations and a novel vector clock compression scheme that are readily applicable to other partial order based analyses.
    Our evaluation with several microbenchmarks and benchmarks shows that \approach has better data race coverage compared to prior techniques at practical run-time overheads.

\end{abstract}


}{}

\iftoggle{asplos}{
    \date{}}{}

\maketitle

\iftoggle{ieeeFormat}{

}{}

\iftoggle{asplos}{
    \thispagestyle{empty}
    \begin{abstract}
        
    \end{abstract}
}{}

\section{Introduction}

\swarnendu{Switch the usage of sound and precise to sound and complete.}




Recent GPU architectures have evolved from supporting bulk-synchronous applications to allowing
fine-grained inter-thread
communication~\cite{ptx-memory-model-nvidia,scord-isca-2020,iguard-sosp-21}. Current synchronization
mechanisms in CUDA are arguably more varied and complicated to use correctly than synchronization on
CPUs. Programmers use involved 
mechanisms composed of barriers, fences, and atomic operations, augmented with scope qualifiers, to protect concurrent accesses to shared variables~\cite{cuda-programming-guide}.
The 
rapidly evolving GPU architecture and the involved synchronization schemes in GPUs introduce
concurrency challenges in writing correct but efficient programs. Concurrency errors, such as data
races, give rise to undesired nondeterminism that can lead to incorrect output or program crashes
and make debugging
difficult~\cite{conc-bug-study-2008,microsoft-exploratory-survey,adversarial-memory,portend-asplos12}. A \emph{data
  race} in a GPU kernel involves concurrent accesses to the same global or shared GPU memory location
with incorrect synchronization, such that both the accesses are not atomic\footnote{Atomic
  operations on shared memory accesses do not guarantee atomicity to regular stores to
  the same address.}, and at least one access is a write.
Several techniques have been proposed to
detect data races in shared and global GPU memory~\cite{curd-pldi-2018,barracuda-pldi-2017,gmrace-tpds-2014,grace-ppopp-2011,scord-isca-2020,ld-taco-2017,ldetector-wodet-2014,haccrg-icpp-2013,iguard-sosp-21}.

\later{
  \swarnendu{Can we give real-world examples to highlight the bad effects of data races in GPU kernels?}
}


\paragraph*{The problem}




Data races are hard to reproduce and debug as they may only manifest non-deterministically with
specific thread interleavings and inputs. Therefore, it is desirable to detect as many data races as
possible by observing only a few (ideally one) executions. However, rapidly evolving GPU
  architecture and programming models complicate reasoning about potential concurrency bugs, tripping up even state-of-the-art dynamic race detectors~\cite{scord-isca-2020,barracuda-pldi-2017,curd-pldi-2018}.

There has been a spurt of work on \emph{predictive} data race detection for CPU
programs~\cite{vindicator-pldi-2018,smarttrack,wcp-pldi-2017,cp-popl-2012,shb-oopsla-2018,sync-preserving-races,depaware-oopsla-2019,rvpredict-pldi-2014,maximal-data-race}.
Predictive techniques observe one dynamic program execution and use partial order relations to
reason about valid alternate ordering among accesses to shared variables. A \emph{predictable} race
manifests when the events during the execution of a program can be reordered to make unsynchronized
memory accesses to shared data variables happen next to each other. Dynamic predictive techniques
are useful because they can scale to large programs and are sound (\ie, reports true data races).
Techniques that track the happens-before (HB) relation have limited predictability, since the number
of races detected is impacted by spurious ordering among concurrent events.
Most existing dynamic data race detectors for GPUs are not
predictive~\cite{boyer-stmcs-2008,grace-ppopp-2011,gmrace-tpds-2014,barracuda-pldi-2017,curd-pldi-2018,haccrg-icpp-2013,ld-taco-2017,ldetector-wodet-2014,iguard-sosp-21},
\ie, they do not reason about data races in other possible interleavings. The \scord race
detector~\cite{scord-isca-2020} uses lockset analysis~\cite{eraser} and therefore has limited
predictive capabilities but misses some classes of predictable data races (\Cref*{sec:predictive}).




\paragraph*{Our approach}


This work explores predictive dynamic race detection for GPUs.
Predictive partial order relations for GPU kernels need to encode the programming and execution model semantics, which a direct port of CPU partial orders will fail to do.
This paper proposes a predictive partial order relation, \emph{\gwcp}, for race detection on GPUs.
\GWCP is based on a sound (has no false positives) and complete (finds all races given constraints) partial order relation called weak-causally-precedes (WCP)~\cite{wcp-pldi-2017} proposed for multithreaded CPU programs.
\gwcp builds on WCP to correctly account for modern GPU capabilities and execution semantics.

We implement \emph{\approach}, a dynamic analysis that tracks the \gwcp relation using vector clocks for CUDA programs.
Using vector clocks for GPU kernels can lead to prohibitive run time and memory overheads~\cite{barracuda-pldi-2017}.
The proposed version of \approach includes several key optimizations to help keep the overheads low in commonly occuring patterns by exploiting redundancies in per-thread vector clocks.
\Approach includes a flexible vector clock compression scheme that is effective in the presence of new scheduling features on \nvidia GPUs~\cite{volta-whitepaper}.
We evaluate \approach against state-of-the-art dynamic data race detectors for GPUs like \barracuda~\cite{barracuda-pldi-2017}, \scord~\cite{scord-isca-2020}, and \iguard~\cite{iguard-sosp-21}. Our evaluation shows that while prior work can both miss data races and raise false alarms, \approach provides better data race coverage with no false positives. This work is the first to explore predictive data race detection for GPU programs to the best of our knowledge.



\paragraph*{Contributions}

This paper makes the following contributions.

\begin{itemize}
  \item it explores predictive partial orders for high-coverage data race detection on GPUs
        \begin{itemize}
          \item we show that existing race detectors either miss or report false data races with several examples;
          \item the proposed \gwcp relation correctly accounts for the GPU thread hierarchy and different synchronization mechanisms to report true races;
        \end{itemize}
  \item a general vector clock compression scheme for per-thread vector clocks exploiting thread hierarchy redundancies and applicable to all vector clock based analyses
  \item comparison of \approach with state-of-the-art techniques to show improved race coverage at practical overheads
  \item an implementation of \barracuda that has been upgraded to deal with modern semantics such as warp-level barriers, intra warp races and scoped atomics
\end{itemize}



\section{Background and Related Work}
\label{sec:background}

The presentation assumes \NVIDIA GPUs and the CUDA programming model. However, the ideas presented
should work with other GPU programming models like OpenCL~\cite{opencl}.

\begin{figure}[t]
  \centering
  \begin{lstlisting}[style=CUDAStyle]
  __global__ void kernel(int *data) {
      data[threadIdx.x] = 2;
      data[1 - threadIdx.x] = 1;
  }
  int main(void) {
      int* d_data;
      cudaMalloc(&d_data, 2 * sizeof(int));
      kernel<<<1,2>>>(d_data);
      return 0;
  }
        \end{lstlisting}
  \caption{A race exposed only with ITS. Lines 2 and 3 do not race with lockstep execution.}
  \label{fig:its}
\end{figure}

\subsection{CUDA Programming Model}
\label{sec:cuda-model}


Compute Unified Device Architecture (CUDA) is a parallel programming model to accelerate GPU programs (\emph{kernels}) on \nvidia GPUs~\cite{cuda-programming-guide}. A CPU (\emph{host}) driver program allocates resources, specifies the hierarchy of threads to be used (\emph{grid}) and launches the kernel on the GPU.
A grid consists of a 1/2/3-dimensional collection of \emph{thread blocks}, and a block is a 1/2/3-dimensional collection of CUDA threads.
A \emph{warp} is the unit of execution on \nvidia GPUs, typically 32 CUDA threads form a warp.
All the threads in a warp execute the same program statement in lockstep on the SIMD hardware for pre-Volta architectures.
From Volta onward, every thread within a warp has its own Program Counter (PC) and call stacks, relaxing the lockstep rule from older architectures, a feature known as \emph{Independent Thread Scheduling} (ITS)~\cite{volta-whitepaper}.


CUDA supports barriers (\eg, \code{\_\_syncthreads()}), atomic read-modify-write instructions (\eg, \code{atomicCAS()}), and memory fence instructions (\eg, \code{\_\_threadfence()}) for synchronization.
For atomics and fences, CUDA exposes three scope qualifiers: \pcode{block}, \pcode{device}, and \pcode{system}, to limit data communication to a subset of relevant threads for better performance.
CUDA does not provide device-wide barriers and provides no standard method for interblock synchronization.
CUDA also does not yet expose lock APIs for synchronization,
although there are \pcode{acquire}/\pcode{release} PTX instructions~\cite{cuda-programming-guide}.
Therefore, programmers often implement ad-hoc lock operations in CUDA: an
atomic compare-and-swap followed by a fence is considered a lock acquire, and an atomic exchange
preceded by a fence is considered a lock release.

\mayant{Define fences when they're used later in the text}

\begin{figure}[t]
  \centering
  \begin{lstlisting}[style=CUDAStyle]
  __global__ void intrawarp(unsigned *data) {
      data[0] = threadIdx.x;
    }
  __global__ void interblk_scope(unsigned *data) {
      atomicExch_block(data, threadIdx.x);
  }
        \end{lstlisting}
  \caption{Kernel \code{intrawarp} has races with multiple threads per block.
    Kernel \code{interblk\_scope} has a data race due to insufficient (\ie, only block-level) scope when invoked with a grid of more than one block.
  }
  \label{fig:race-examples}
\end{figure}



{Intrawarp} races occur when 
threads from the same warp {write} to the same memory location.
While lockstep execution implies two instructions executed by a warp cannot race,
we consider synchronization-free intrawarp communication as a data race due to ITS~\cite{curd-pldi-2018} (\eg, Figure~\ref{fig:its}).
{Interwarp} races arise when the threads are from different warps.
Incorrect usage of scopes with atomics, fences, and locks 
also lead to data races~\cite{scord-isca-2020}.
Figure~\ref{fig:race-examples} shows an intrawarp data race and a race due to insufficient
scope.
We ignore inter-kernel data races in this work.


\subsection{Dynamic Detection of Data Races}

In the following, we discuss state-of-the-art
techniques for dynamic data race detection on GPUs.


\emph{Barracuda}~\cite{barracuda-pldi-2017} 
checks whether concurrent accesses from different threads are separated by the \emph{happens-before} (HB)~\cite{happens-before} relation extended with GPU execution semantics.
The HB relation is a partial order defined over the events in a dynamic kernel trace $\alpha$. Given two events $a, b \in \alpha$ such that $a$ is before $b$ in the trace (denoted by $a <^{\alpha}_{tr} b$), event $a$ happens before (\ie, is ordered with) event $b$ if (i) $a$ and $b$ are performed by the same thread (intra-thread order), (ii) $a$ is part of a warp that executes before $b$'s warp (interwarp order), (iii) $a$ or $b$ is a barrier, 
or a barrier separates $a$ and $b$, or (iv) $a$ and $b$ access the same synchronization variable where $a$ is a release and $b$ is an acquire operation, and both operations are either at the block scope within the same thread block or at least one operation is at the global scope (inter-thread synchronization).
A data race occurs when $a$ and $b$ access the same location, one of the accesses is a write, the operations are not both atomic, and neither $a$ nor $b$ happen before each other.

We refer to the partial order used in \Barracuda as \emph{scoped HB} in future discussions.

Barracuda pushes operations executed on the GPU to a queue shared with the host, and the host
consumes the operations and runs the race detection logic.
\Barracuda tracks intrawarp races by taking the join of all vector clocks currently active in the warp after every instruction.
\barracuda
considers scopes in only fence operations and ignores scopes in other synchronization operations
such as atomics and locks.

\emph{CURD}~\cite{curd-pldi-2018} speeds up race detection 
when the synchronization in the kernel only involves barriers. CURD identifies such kernels using static analysis and uses compiler instrumentation to track the read and write accesses in synchronization-free regions\footnote{An SFR is a sequence of non-synchronization instructions executed by a thread, delimited by synchronization operations (e.g., atomics and barriers).}. CURD intersects read and write sets \emph{on} the GPU to check for data races.
We focus on \barracuda in this work since CURD does not aim to improve race coverage over \barracuda.



\emph{\Scord}~\cite{scord-isca-2020} uses lockset analysis to identify scoped races induced due to
misuse of lock-based critical sections. Lockset-based algorithms assume a consistent locking
discipline for accessing shared variables~\cite{eraser}, a condition difficult to enforce due to ad-hoc lock implementations. Since lockset analysis cannot detect data
races in kernels that use only atomics or barriers for
synchronization~\cite{grace-ppopp-2011,curd-pldi-2018}, \scord extends the HB mechanism to detect
scoped races due to barriers and atomics. \Scord proposes hardware extensions for efficient race
detection. To keep the hardware overhead bounded, \Scord maintains metadata for every 4~B of global
memory and only stores metadata for recent accesses, leading to both
missed and false races. Furthermore, \Scord
maintains metadata at the warp granularity and does not detect intrawarp races.

In recent work, \emph{\iguard}~\cite{iguard-sosp-21} 
addresses the shortcomings of \scord by using binary instrumentation 
instead of hardware extensions. \iguard achieves good performance by moving all race detection logic
to the GPU and uses Unified Shared Memory for maintaining metadata.
Besides supporting ITS, the core analysis in \iguard is largely similar to \scord.

The ITS mechanism necessitates a thread-level analysis rather than a warp-level analysis.
Both \barracuda and \scord fail to detect the race in \Cref*{fig:its}; \barracuda models
lockstep execution within a warp and \scord maintains metadata at warp granularity.
ITS also makes it challenging to design metadata compression schemes~\cite{barracuda-pldi-2017} which rely on per-thread vector clock entries being the same for \emph{all} threads in a warp as individual threads can now synchronize with each other.

\section{Predictive Data Race Detection}
\label{sec:predictive}


\begin{figure}[t]
	\centering
	\begin{minipage}[t]{0.15\linewidth}
		\textbf{Event}
		\begin{lstlisting}[style=TraceStyle]
e1
e2
e3
e4
e5
e6
e7
e8
		\end{lstlisting}
	\end{minipage}%
	\begin{minipage}[t]{0.35\linewidth}
		\textbf{Thread 1}
		\begin{lstlisting}[style=TraceStyle]
wr(x)
acq(l)
wr(y)
rel(l)


		\end{lstlisting}
	\end{minipage}%
	\begin{minipage}[t]{0.35\linewidth}
		\textbf{Thread 2}
		\begin{lstlisting}[style=TraceStyle]




acq(l)
wr(x)
wr(y)
rel(l)
		\end{lstlisting}
	\end{minipage}
	\swarnendu{We should adjust our thread headings keeping in mind CUDA.}
	\caption{An example of a predictable race detected by WCP but not by HB (reproduced from the WCP paper~\cite{wcp-pldi-2017}).}
	\label{fig:no-CP}
\end{figure}

Given the non-deterministic nature of reproducing data races and the high cost in subsequent
debugging~\cite{conc-bug-study-2008,microsoft-exploratory-survey,portend-asplos12}, researchers have
explored ways to improve race detection coverage for CPU programs
(\ie, detect as many true races as possible).
A few techniques randomize the thread scheduler and perturb the execution
to explore different valid interleavings to maximize detecting races that can occur across schedules~\cite{randomized-scheduler,racageddon,racefuzzer}.
Predictive race detection techniques observe one dynamic program execution and aim to detect data
races that can occur in other
\emph{correct} reorderings of memory accesses.
A necessary criterion to ensure correct reorderings of a trace is to maintain write-to-read orders so that the new interleaving explores the same code branches~\cite{smarttrack,vindicator-pldi-2018}.
Predictive techniques for CPU programs use partial
order relations that are weaker than HB to reconstruct valid memory reorderings and have shown
promise in detecting more data races in real-world
applications~\cite{vindicator-pldi-2018,wcp-pldi-2017,smarttrack,cp-popl-2012,maximal-data-race,rvpredict-pldi-2014,dighr,pavlogiannis-2019,depaware-oopsla-2019,sync-preserving-races}.



\begin{figure}[t]
    \centering
    \begin{minipage}[t]{0.15\linewidth}
        \textbf{Event}
        \begin{lstlisting}[style=TraceStyle]
e1
e2
e3
        \end{lstlisting}
    \end{minipage}%
    \begin{minipage}[t]{0.35\linewidth}
        \textbf{Thread 1}
        \begin{lstlisting}[style=TraceStyle]
wr(x)
threadfence

        \end{lstlisting}
    \end{minipage}%
    \begin{minipage}[t]{0.35\linewidth}
        \textbf{Thread 2}
        \begin{lstlisting}[style=TraceStyle]


wr(x)
        \end{lstlisting}
    \end{minipage}
    \caption{The interleaving is not racy if the accesses are from the same block, but indicates a predictable race missed by \scord.}
    \label{fig:predictable-race-scord}
\end{figure}

\subsection{Predictive Power of Prior Work}

\swarnendu{Compare with iGUARD~\cite{iguard-sosp-21}.}

We now discuss the limitations of \barracuda and \scord for predictive race detection of CUDA kernels.
Consider the interleaving shown in Figure~\ref{fig:no-CP}, where the composition of synchronization order (via the sequence of release and acquire calls) and per-thread program order (
\pcode{\small{e1$\to$e4$\to$e5$\to$e6}})
establish an ordering between the events \pcode{e1} and \pcode{e6}. Techniques like \Barracuda that
use the HB relation will not report the predictable race between \pcode{e1} and \pcode{e6}.
The dynamic interleaving 
of instructions during an execution does not impact the race coverage of lockset algorithms. Lockset
analysis reports a data race since the lockset at \pcode{e6} contains the lock \pcode{l}, whereas
the lockset at \pcode{e1} is empty. This allows \scord to detect the race successfully. To avoid
false positives, \Scord omits lockset analysis if no locks were held during the previous and current
memory accesses, which leads to \Scord missing predictable races.
In Figure \ref{fig:predictable-race-scord}, both accesses are unprotected. Therefore, \scord
performs only fence-based analysis, which cannot detect the potential race on \pcode{x}. We verified that on \emph{forcing} the racy interleaving through delays,
\scord can detect the race. We have also verified that using a lock to protect one of the accesses
but not the other will also allow \scord to detect the potential race via locksets.

\subsection{Predictive Partial Orders}
\label{sec:design-overview}


Several partial order relations weaker than HB have been proposed for predictive data race detection
on CPUs (\eg, causally-precedes (CP)~\cite{cp-popl-2012}, weak-causally-precedes
(WCP)~\cite{wcp-pldi-2017}, schedulable-happens-before (SHB)~\cite{shb-oopsla-2018},
strong-dependently-precedes (SDP)~\cite{depaware-oopsla-2019}, doesn't-commute
(DC)~\cite{vindicator-pldi-2018}, M2~\cite{pavlogiannis-2019}, weak-doesn't-commute
(WDC)~\cite{smarttrack}), and sync-preserving races (SyncP)~\cite{sync-preserving-races}. This work uses the WCP relation as the baseline and extends it to the GPU
programming model.
We ignore other partial order relations like CP and M2 since they are expensive to
track~\cite{raptor-arxiv}.
Recent work has shown the SHB relation to be imprecise~\cite{sync-preserving-races}.
Even though DC and WDC can potentially find more data races than WCP, the relations can report false positives and require additional ``vindication'' analysis to prune the false races.
Recent work on sync-preserving-races~\cite{sync-preserving-races} show interleavings where WCP may
miss reporting predictable races. However, WCP can re-order critical sections while SyncP cannot and
hence miss races, which can be a limitation in GPU kernels with massive parallelism.

\paragraph*{WCP}

The key insight in WCP~\cite{wcp-pldi-2017} is that release-to-acquire ordering of conflicting
critical sections is conservative and can be further relaxed based on the order of events
\emph{within} critical sections. WCP only orders the release of the first critical section to the
conflicting event of the second critical section.
Given a trace $\alpha$ of events in a multithreaded execution, $<_{\textrm{WCP}}^{\alpha}$ is
the smallest relation that satisfies the following conditions.





\begin{enumerate}[(i)]
      \item For a 
            release event $r$ on lock $l$ and a read/write event $e$ on memory location $x$ in a critical section on the same lock with $r <^{\alpha}_{tr} e$ (\ie, $r$ is ordered before $e$ in $\alpha$), if the critical section of $r$ contains an event conflicting with $e$, then $r$ is WCP-ordered with $e$.
      \item For two release events $r_1$ and $r_2$ on lock $l$ with $r_1 <^{\alpha}_{tr} r_2$, if the critical sections corresponding to $r_{1}$ and $r_{2}$ contain WCP-ordered events, then $r_1$ is ordered before $r_2$ via WCP.
      \item WCP composes with the happens-before (HB) order. That is,
            $<_{\textrm{WCP}}^{\alpha} = (<_{\textrm{WCP}}^{\alpha} \circ \leq_{\textrm{HB}}^{\alpha}) = (\leq_{\textrm{HB}}^{\alpha} \circ <_{\textrm{WCP}}^{\alpha})$.
\end{enumerate}

The relation $\leq^{\alpha}_{\textrm{WCP}} = (<_{\textrm{WCP}}^{\alpha} \cup \leq_{\textrm{TO}}^{\alpha})$, where $\leq_{\textrm{TO}}^{\alpha}$ is the program order, is defined to be the WCP partial order.


Figure~\ref{fig:no-CP} shows a predictable race on \code{x}, via the interleaving \pcode{\small{e5$\to$e1$\to$e6}}, that is detected by WCP.
However, HB orders releases to acquires on the same lock and composes with program order leading to
\pcode{\small{e1$\leq_{\textrm{TO}}$e4$\leq_{\textrm{HB}}$e5$\leq_{\textrm{TO}}$e6}}, hiding the race.
With WCP, only the release operation \pcode{e4} is ordered before the conflicting access at \pcode{e7} by rule (i).





However, the WCP relation always reports true data races (or results in a deadlock),
which is a desirable trait not to waste developer time and effort~\cite{literace}. Furthermore, WCP
can also be efficiently implemented compared to other partial order relations. These properties make
WCP an attractive candidate for implementing a predictive race analysis for GPU kernels.



\section{Extending WCP to GPU Programs}


Directly applying predictive partial orders for CPUs, such as WCP, does not work on GPUs.
In the following, we discuss the challenges in encoding GPU synchronization semantics in a partial order. 
We ignore modeling lockstep execution since it is not guaranteed in newer GPU architectures.

\paragraph*{Barrier synchronization}


Kernels may use scoped barriers to synchronize threads~\cite{curd-pldi-2018}, and memory accesses separated by a barrier can never race. A predictive partial order must explicitly order such accesses to avoid false positives.




\paragraph*{Scoped atomics}


Atomic operations on a CPU do not race, but atomic operations that are insufficiently scoped \emph{can} race on a GPU.
A predictive partial order for GPUs should encode the notion of scopes for atomics
and should not order two insufficiently scoped atomic accesses to a memory location.

\paragraph*{Scoped locks}


Scoped atomics
or fences constitute \emph{scoped} locks. Two locks \emph{overlap} in scope if either
one of them is device-scoped or threads from the same block hold both the locks. Mutual exclusion is
not guaranteed for non-overlapping critical sections. A predictive partial order for GPUs should not
order non-overlapping critical sections, even if they contain conflicting accesses.

\swarnendu{We should associate the above cases to examples, especially scoped locks.}
\mayant{Consider removing the discussion on differently scoped acquire and releases}



\begin{definition}
  Given a trace $\alpha$, $<_{\textrm{\GWCP}}^{\alpha}$ is the smallest relation that satisfies the following conditions.

  \begin{enumerate}[(i)]
    \item For a release event $r$ on lock $l$ and a read/write event $e$ on memory location $x$ in a critical section on the same lock with $r <^{\alpha}_{tr} e$, if the critical section of $r$ contains an event conflicting with $e$ and the scopes in these critical sections overlap, then $r$ is ordered before $e$ by \GWCP.
    \item For two release events $r_1$ and $r_2$ on lock $l$ with $r_1 <^{\alpha}_{tr} r_2$, if the critical sections corresponding to $r_{1}$ and $r_{2}$ contain \GWCP-ordered events and the scopes of these releases overlap, then $r_1$ is ordered before $r_2$ by \GWCP.
    \item For a scoped barrier $b$ covering threads $t_1$ and $t_2$, if $e_i$ and $e_i'$ are events on $t_i$ ($i\in\left[1,2\right]$) such that $e_i <^{\alpha}_{tr} b <^{\alpha}_{tr} e_i'$, then $e_1 <_{\textrm{\GWCP}}^{\alpha} e_2'$ and $e_2 <_{\textrm{\GWCP}}^{\alpha} e_1'$.
    \item \GWCP composes with scoped HB order:
          $<_{\textrm{\GWCP}}^{\alpha} = (<_{\textrm{\GWCP}}^{\alpha} \circ \leq_{\textrm{HB}}^{\alpha}) = (\leq_{\textrm{HB}}^{\alpha} \circ <_{\textrm{\GWCP}}^{\alpha})$.
  \end{enumerate}

  The relation $\leq^{\alpha}_{\textrm{\GWCP}} = (<_{\textrm{\GWCP}}^{\alpha} \cup \leq_{\textrm{TO}}^{\alpha})$ is defined to be the \emph{weak-causally-precedes for GPU} partial order.
\end{definition}

\sagnik{Should mention how we use the scoped version of HB too. Actually, currently we don't in the code but we should. Noting here as a TODO.}
\swarnendu{How are the scoped atomics encoded in \gwcp? I remember we had decided to add a few extra conditions.}

\GWCP 
respects the above requirements for ordering concurrent events on
GPUs. A race is \GWCP-predictable if it is exposed in a correctly reordered trace with
respect to \GWCP. \approach is sound and precise over \GWCP-predictable races.

\later{\swarnendu{Do we need to provide some intuition for the why \gwcp respects the above requirements?}}

\paragraph*{Example}


Figure~\ref{fig:predictable-race-gwcp} shows a predictable race on \pcode{e1} and \pcode{e14} when \thr{T2} and \thr{T3} are from different blocks. Event \pcode{e4} is ordered before \pcode{e6} via rule (i), but note that \pcode{e10} and \pcode{e11} will not be ordered by \GWCP despite being conflicting since they are of insufficient scope. Therefore, there is no ordering between \pcode{e1} and \pcode{e14} and the race is detected by \gwcp. \barracuda fails to detect this race because it orders acquires to releases leading to the ordering \pcode{\small{e1$\to$e4$\to$e5$\to$e10$\to$e11$\to$e14}}.
\scord is unable to catch the race because it omits lockset detection for empty locksets.
The use of a fence instruction by the acquire at \pcode{e2} leads \scord to not trigger a missing-fence-related race.


\begin{figure}
    \centering
    \begin{minipage}[t]{0.15\linewidth}
        \textbf{Event}
        \begin{lstlisting}[style=TraceStyle]
e1
e2
e3
e4
e5
e6
e7
e8
e9
e10
e11
e12
e13
e14
            \end{lstlisting}
    \end{minipage}%
    \begin{minipage}[t]{0.28\linewidth}
        \textbf{Blk 1,Thr 1}
        \begin{lstlisting}[style=TraceStyle]
wr(x)
acq(m)
wr(y)
rel(m)








        \end{lstlisting}
    \end{minipage}%
    \begin{minipage}[t]{0.28\linewidth}
        \textbf{Blk 1,Thr 2}
        \begin{lstlisting}[style=TraceStyle]




acq(m)
wr(y)
rel(m)
acqblk(n)
wr(z)
relblk(n)






        \end{lstlisting}
    \end{minipage}%
    \begin{minipage}[t]{0.28\linewidth}
        \textbf{Blk 2,Thr 3}
        \begin{lstlisting}[style=TraceStyle]










acqblk(n)
wr(z)
relblk(n)
wr(x)
        \end{lstlisting}
    \end{minipage}
    \caption{Predictable race detected by \GWCP.}
    \label{fig:predictable-race-gwcp}
\end{figure}

\subsection{Tracking \GWCP}
\label{sec:algorithm}

Algorithms~\ref{algo:gwcp-acm-sync} and \ref{algo:gwcp-acm-mem} show the rules to track the \gwcp relation via an on-the-fly dynamic analysis.
We use the same notation used in the WCP work~\cite{wcp-pldi-2017}.

\paragraph*{Metadata}

The analysis maintains the following metadata.
\begin{itemize}
  \item Per-thread local time $\mathbb{N}_t$ and vector clocks $\mathbb{P}_t$ and $\mathbb{H}_t$,
  \item Per-lock vector clocks $\mathbb{P}_l$ and $\mathbb{H}_l$ and read and write sets $R_l$ and $W_l$,
  \item Per-location read and write vector clocks $R_x$ and $W_x$,
  \item Per-thread locksets $L_t$,
  \item Acquire and release event queues \pcode{Acq}$_l$(t) and \pcode{Rel}$_l$(t),
  \item Vector clocks $\mathbb{L}_{\langle l,s \rangle, x}^r$ and $\mathbb{L}_{\langle l,s \rangle, x}^w$ per combination of $\langle$memory location, lock$\rangle$ pair, where $l$ represents the location and $s$ represents the scope.
\end{itemize}


$\mathbb{P}_t$ refers to the per-thread vector clock (PTVC) corresponding to the relation $<_{\textrm{\GWCP}}^{\alpha}$. $\mathbb{P}_l$ is the corresponding lock vector clock.
$\mathbb{N}_t$ is combined with $\mathbb{P}_t$ to form $\mathbb{C}_t$ which represents the actual
logical time used to capture the \GWCP relation $\leq^{\alpha}_{\textrm{\GWCP}}$. $\mathbb{H}_t$ and
$\mathbb{H}_l$ are per-thread and lock vector clocks for HB ordering. The happens-before ordering
rules are similar to that used by \barracuda. $L_t$ is the lockset for the thread $t$, and tracks
the set of locks currently held by $t$ along with the lock scopes. \pcode{Acq}$_l$ and
\pcode{Rel}$_l$ queues store acquire and release events on lock variables along with the scopes.


$R_l$ and $W_l$ are read and write sets for critical sections on $l$. $R_x$ and $W_x$ represent vector clocks to track the last reader/writer thread(s) and the corresponding access times for a memory location $x$.
$W_x$ is an epoch, which is a pair of the local time of the last accessing thread and its ID~\cite{fasttrack}.
$R_x$ and $W_x$ are used to check for ordering between accesses, and are updated with $\mathbb{C}_t$ on each access.


The lock identifier in certain data structures has been expanded from a memory value to a pair to
allow \GWCP to take scopes into account. The algorithm stores the scope with which a lock has been
acquired.
We omit the updates to $R_l$, $W_l$, and $L_t$ in the pseudocode to save space. $L_t$ is the per thread lockset and is maintained by inserting a lock into the set on aquires and removing it on a release. $R_l$ and $W_l$ are the read set and write sets for a particular thread respectively. These are updated on reads and writes by checking the current lockset and inserting into the appropriate sets.
On a release, $R_l$ and $W_l$ are reset.


\swarnendu{SD/MM: I have skipped the following text, please revise it and let me know.}

In Procedure \code{sync}, the parameter \code{scope} is a set of TIDs that are part of the scope of the current synchronization operation \code{(\_\_syncthreads() or \_\_syncwarp())}. $t$ is the parameter to specify the calling thread. In the \code{acquire} and \code{release} procedures, $l$ parameter is the lock variable accessed and the $s$ parameter is the aforementioned scope that specifies the scope with which the current lock variable was used.

\mayant{Go over the last section's example, but in the algorithm's context now}

\iftoggle{acmFormat}{
  \begin{small}
    \begin{algorithm}[t]
        \caption{Tracking \GWCP for sync operations 
        }
        \label{algo:gwcp-acm-sync}
        \begin{algorithmic}[1]

            \Procedure{sync}{$scope$}
            \State $joined \coloneqq joined_{HB} \coloneqq \phi$

            \ForAll {$t \in scope$}
            \State $\mathbb{C}_t \coloneqq \mathbb{P}_t\{t\coloneqq \mathbb{N}_t\}$
            \State $joined \coloneqq joined \sqcup \mathbb{C}_t$
            \State $joined_{HB} \coloneqq joined_{HB} \sqcup \mathbb{H}_t$
            \EndFor

            \ForAll {$t \in scope$}
            \State $\mathbb{P}_t \coloneqq joined$
            \State $\mathbb{H}_t \coloneqq joined_{HB}$
            \EndFor

            \State $\mathbb{N}_t \coloneqq \mathbb{N}_t$ + 1
            \State $\mathbb{H}_t \coloneqq \mathbb{H}_t$[$t \coloneqq$ $\mathbb{H}_t$($t$) + 1]
            \EndProcedure


            \Procedure{acquire}{$t$, $l$, $s$}
            \State $\mathbb{P}_t \coloneqq \mathbb{P}_t \sqcup \mathbb{P}_{\langle l,s \rangle}$
            \State $\mathbb{H}_t \coloneqq \mathbb{H}_t \sqcup \mathbb{H}_{\langle l,s \rangle}$
            \If {$s$ = \textsl{DEVICE}} \Comment{Device-level lock}
            \ForAll {$blk \in grid$}
            \State $\mathbb{P}_t \coloneqq \mathbb{P}_t \sqcup \mathbb{P}_{\langle l,\mathtt{blk} \rangle}$ \Comment{$\mathtt{blk}$ is a block-level lock}
            \State $\mathbb{H}_t \coloneqq \mathbb{H}_t \sqcup \mathbb{H}_{\langle l,\mathtt{blk} \rangle}$
            \EndFor
            \Else
            \State $\mathbb{P}_t \coloneqq \mathbb{P}_t \sqcup \mathbb{P}_{\langle l,\mathtt{DEVICE}\rangle}$
            \State $\mathbb{H}_t \coloneqq \mathbb{H}_t \sqcup \mathbb{H}_{\langle l,\mathtt{DEVICE}\rangle}$
            \EndIf
            \ForAll {$t' \neq t$}
            \State Acq$_l$($t'$).enque$\left(\langle\mathbb{C}_t, s\rangle\right)$
            \EndFor
            \EndProcedure


            \Procedure{release}{$t$, $l$, $R$, $W$, $s$}
            \While{Acq$_l\left(t\right)$.front() $\sqsubseteq \mathbb{C}_t$}
            \State Acq$_l$($t$).deque()
            \State $\langle$VC, $scope \rangle$ $\leftarrow$ Rel$_l$($t$).deque()

            \If {$scope$ = \textsl{DEVICE} \Or  $scope$ = $\mathtt{blk}$}
            \State $\mathbb{P}_t \coloneqq \mathbb{P}_t \sqcup$ $VC$
            \EndIf
            \EndWhile

            \ForAll {$x \in R$}
            \State $\mathbb{L}_{\langle l,s \rangle, x}^r \coloneqq \mathbb{L}_{\langle l,s \rangle, x}^r \sqcup \mathbb{H}_t$
            \EndFor

            \ForAll {$x \in W$}
            \State $\mathbb{L}_{\langle l,s \rangle, x}^w \coloneqq \mathbb{L}_{\langle l,s \rangle, x}^w \sqcup \mathbb{H}_t$
            \EndFor

            \State $\mathbb{H}_{\langle l,s \rangle} \coloneqq \mathbb{H}_t$
            \State $\mathbb{P}_{\langle l,s \rangle} \coloneqq \mathbb{P}_t$

            \ForAll {$t' \neq t$}
            \State Rel$_l$($t'$).enque($\langle \mathbb{H}_t,s \rangle$)
            \EndFor

            \State $\mathbb{N}_t \coloneqq \mathbb{N}_t$ + 1
            \State $\mathbb{H}_t \coloneqq \mathbb{H}_t$[$t \coloneqq$ $\mathbb{H}_t$($t$) + 1]
            \EndProcedure
        \end{algorithmic}
    \end{algorithm}
\end{small}

\swarnendu{Algorithm~\ref{algo:gwcp-acm} Fix line 10, scope=0 has no meaning. Line 21: What is s? Add a few comments to explain the steps.}

  \begin{small}
    \begin{algorithm}[t]
        \caption{Tracking \GWCP for memory accesses}
        \label{algo:gwcp-acm-mem}
        \begin{algorithmic}[1]

            \Procedure{read}{$t$, $x$, $L$}
            \ForAll {$\langle l, s \rangle \in L$}
            \State $\mathbb{P}_t \coloneqq \mathbb{P}_t \sqcup \mathbb{L}_{\langle l,s\rangle, x}^w$
            \If {$s$ = \textsl{DEVICE}}
            \ForAll {$blk \in grid$}
            \State $\mathbb{P}_t \coloneqq \mathbb{P}_t \sqcup \mathbb{L}_{\langle l, blk \rangle, x}^w$
            \EndFor
            \Else
            \State $\mathbb{P}_t \coloneqq \mathbb{P}_t \sqcup \mathbb{L}_{\langle l, \mathtt{DEVICE}\rangle, x}^w$
            \EndIf
            \EndFor
            \EndProcedure


            \Procedure{write}{$t$, $x$, $L$}
            \ForAll {$\langle l, s \rangle \in L$}
            \State $\mathbb{P}_t \coloneqq \mathbb{P}_t \sqcup \mathbb{L}_{\langle l,s \rangle, x}^w$
            \State $\mathbb{P}_t \coloneqq \mathbb{P}_t \sqcup \mathbb{L}_{\langle l,s \rangle, x}^r$
            \If {$s$ = \textsl{DEVICE}}
            \ForAll {$blk \in grid$}
            \State $\mathbb{P}_t \coloneqq \mathbb{P}_t \sqcup \mathbb{L}_{\langle l, blk \rangle, x}^w$
            \State $\mathbb{P}_t \coloneqq \mathbb{P}_t \sqcup \mathbb{L}_{\langle l, blk \rangle, x}^r$
            \EndFor
            \Else
            \State $\mathbb{P}_t \coloneqq \mathbb{P}_t \sqcup \mathbb{L}_{\langle l, DEVICE \rangle, x}^w$
            \State $\mathbb{P}_t \coloneqq \mathbb{P}_t \sqcup \mathbb{L}_{\langle l, DEVICE \rangle, x}^r$
            \EndIf
            \EndFor
            \EndProcedure
        \end{algorithmic}
    \end{algorithm}
\end{small}

\swarnendu{Add a few short comments to explain the steps.}

}{}

\iftoggle{ieeeFormat}{
  \begin{algorithm}[t]
    \caption{SYNC(scope)}
    \label{algo:gwcp-sync}
    \begin{algorithmic}[1]
        \STATE $joined \leftarrow VC()$
        \STATE $joined_{HB} \leftarrow VC()$

        \FORALL {$t \in$ scope}
        \STATE $\mathbb{C}_t \coloneqq \mathbb{P}_t\{t\coloneqq \mathbb{N}_t\}$
        \STATE $joined \coloneqq joined \sqcup \mathbb{C}_t$
        \STATE $joined_{HB} \coloneqq joined_{HB} \sqcup \mathbb{H}_t$
        \ENDFOR

        \FORALL {$t \in$ scope}
        \STATE $\mathbb{P}_t \coloneqq joined$
        \STATE $\mathbb{H}_t \coloneqq joined_{HB}$
        \ENDFOR

        \STATE $\mathbb{N}_t \coloneqq \mathbb{N}_t + 1$
        \STATE $increment(\mathbb{H}_t)$
    \end{algorithmic}
\end{algorithm}

\begin{algorithm}
    \label{algo:gwcp-acq}
    \caption{ACQUIRE($t, l, s$)}
    \begin{algorithmic}[1]
        \STATE $\mathbb{P}_t \coloneqq \mathbb{P}_t \sqcup \mathbb{P}_{<l,s>}$
        \STATE $\mathbb{H}_t \coloneqq \mathbb{H}_t \sqcup \mathbb{H}_{<l,s>}$
        \IF {$s$ = 0}
        \FORALL {$block \in grid$}
        \STATE $\mathbb{P}_t \coloneqq \mathbb{P}_t \sqcup \mathbb{P}_{<l,block+1>}$
        \STATE $\mathbb{H}_t \coloneqq \mathbb{H}_t \sqcup \mathbb{H}_{<l,block+1>}$
        \ENDFOR
        \ELSE
        \STATE $\mathbb{P}_t \coloneqq \mathbb{P}_t \sqcup \mathbb{P}_{<l,0>}$
        \STATE $\mathbb{H}_t \coloneqq \mathbb{H}_t \sqcup \mathbb{H}_{<l,0>}$
        \ENDIF
        \FORALL {$t' \neq t$}
        \STATE Acq$_l(t')$.enque$(<\mathbb{C}_t, s>)$
        \ENDFOR
    \end{algorithmic}
\end{algorithm}

\begin{algorithm}
    \caption{RELEASE($t, l, R, W, s$)}
    \label{algo:gwcp-release}
    \begin{algorithmic}[1]
        \WHILE{Acq$_l(t)$.front$()$ $\sqsubseteq \mathbb{C}_t$}
        \STATE Acq$_l(t)$.deque()
        \STATE <VC, scope> $\leftarrow$ Rel$_l(t)$.deque()

        \IF {$scope$ = 0 \OR  $scope$ = $block(t)$ + 1}
        \STATE $\mathbb{P}_t \coloneqq \mathbb{P}_t \sqcup$ $VC$
        \ENDIF
        \ENDWHILE
        \FORALL {$x \in R$}
        \STATE $\mathbb{L}_{<l,s>, x}^r \coloneqq \mathbb{L}_{<l,s>, x}^r \sqcup \mathbb{H}_t$
        \ENDFOR
        \FORALL {$x \in W$}
        \STATE $\mathbb{L}_{<l,s>, x}^w \coloneqq \mathbb{L}_{<l,s>, x}^w \sqcup \mathbb{H}_t$
        \ENDFOR
        \STATE $\mathbb{H}_{<l,s>} \coloneqq \mathbb{H}_t$
        \STATE $\mathbb{P}_{<l,s>} \coloneqq \mathbb{P}_t$
        \FORALL {$t' \neq t$}
        \STATE Rel$_l(t')$.enque$(<\mathbb{H}_t,s>)$
        \ENDFOR
        \STATE $\mathbb{N}_t \coloneqq \mathbb{N}_t + 1$
        \STATE $increment(\mathbb{H}_t)$
    \end{algorithmic}
\end{algorithm}

\begin{algorithm}
    \caption{READ($t, x, L$)}
    \label{algo:gwcp-read}
    \begin{algorithmic}[1]
        \FORALL {$<l, s> \in L$}
        \STATE $\mathbb{P}_t \coloneqq \mathbb{P}_t \sqcup \mathbb{L}_{<l,s>, x}^w$
        \IF {$s$ = 0}
        \FORALL {$block \in grid$}
        \STATE $\mathbb{P}_t \coloneqq \mathbb{P}_t \sqcup \mathbb{L}_{<l, block+1>, x}^w$
        \ENDFOR
        \ELSE
        \STATE $\mathbb{P}_t \coloneqq \mathbb{P}_t \sqcup \mathbb{L}_{<l, 0>, x}^w$
        \ENDIF
        \ENDFOR
    \end{algorithmic}
\end{algorithm}

\begin{algorithm}
    \caption{WRITE($t, x, L$)}
    \label{algo:gwcp-write}
    \begin{algorithmic}[1]
        \FORALL {$<l, s> \in L$}
        \STATE $\mathbb{P}_t \coloneqq \mathbb{P}_t \sqcup \mathbb{L}_{<l,s>, x}^w$
        \STATE $\mathbb{P}_t \coloneqq \mathbb{P}_t \sqcup \mathbb{L}_{<l,s>, x}^r$
        \IF {$s$ = 0}
        \FORALL {$block \in grid$}
        \STATE $\mathbb{P}_t \coloneqq \mathbb{P}_t \sqcup \mathbb{L}_{<l, block+1>, x}^w$
        \STATE $\mathbb{P}_t \coloneqq \mathbb{P}_t \sqcup \mathbb{L}_{<l, block+1>, x}^r$
        \ENDFOR
        \ELSE
        \STATE $\mathbb{P}_t \coloneqq \mathbb{P}_t \sqcup \mathbb{L}_{<l, 0>, x}^w$
        \STATE $\mathbb{P}_t \coloneqq \mathbb{P}_t \sqcup \mathbb{L}_{<l, 0>, x}^r$
        \ENDIF
        \ENDFOR
    \end{algorithmic}
\end{algorithm}

}{}





The four major differences in GPU synchronization have been addressed as follows:

\paragraph*{Shared memory}
GPUs have a hierarchy in memory that mimics the hierarchy in threads. This leads to two different kinds of memory we monitor: \emph{shared} and \emph{global} memory. While global memory is accessible to the entire GPU device, shared memory by definition, is private to a threadblock and thus, can never be involved in an inter block race. To address this, we promote every memory location to indicate whether it is a global or shared memory and for shared memory we maintain which block it belongs to. This helps us treat shared memory per block as a different memory location.

\paragraph*{Scoped atomics}
We maintain if the last memory access to a location was atomic, along with its scope. Before reporting a race we check whether both accesses are atomics with overlapping scopes, in which case we omit the race report.

\paragraph*{Scoped locks}
Our modification to the WCP algorithm treats differently scoped locks as different locks. On release operations we join PTVCs into the held lock's clock at the appropriate scope, as shown in lines $25-29$ in Algorithm~\ref{algo:gwcp-acm-sync}. On acquire operations, PTVCs are joined with block and/or device scoped lock clocks depending on the scope of the lock. For device level locks, we need joins with both device lock clocks and also with every other block's lock clocks, as shown in lines $16-21$ in Algorithm~\ref{algo:gwcp-acm-sync}. This allows us to have an edge when at least one of the scopes used is sufficient to cover both events.


\paragraph*{Barriers}
At a barrier, all threads involved synchronize with each other. Thus, we take a join of all PTVCs involved in the barrier. This means every thread in the block for a \code{\_\_syncthreads()} and for active threads in the warp for a \code{\_\_syncwarp()} instruction. At this point, the $\mathbb{C}_t$ clocks need to be joined and not just $\mathbb{P}_t$'s since $\mathbb{C}_t$ is the vector clock maintaining actual logical time (line 4 in Algorithm~\ref{algo:gwcp-acm-sync}).


\section{Implementation}
\label{sec:impl}



We implement a prototype tool, \approach, to track \gwcp. \Approach uses the NVBit\footnote{\url{https://github.com/NVlabs/NVBit}} dynamic binary instrumentation framework~\cite{nvbit-micro-2019} from \NVIDIA.
The \barracuda\footnote{\url{https://github.com/upenn-acg/barracuda}} implementation uses a custom
binary instrumentation framework developed in-house. The \scord artifact\footnote{\url{https://github.com/csl-iisc/ScoRD}} uses GPGPU-Sim~\cite{accelsim-isca-2020}, a simulator for \NVIDIA GPUs.
Therefore, we have also reimplemented \barracuda and \scord with NVBit to allow a fair comparison of the techniques. Using binary instrumentation has the advantage of not relying on source code and provides flexibility in place of the closed-source \nvidia toolchain.



The NVBit tools inspect SASS 
instructions and register callbacks 
on memory accesses and synchronization operations. All the implementations instrument the same instructions.
The callbacks 
create event objects 
and push them to a communication queue shared with the host. The host cores consume the events from the channels and run the race detection analysis.
In the following, we discuss optimizations to improve the performance of our implementations.

\paragraph*{Coalesced event processing}

The instrumentation of memory accesses and block and warp synchronization generate only one event per warp. For memory accesses, the event captures the memory accesses of all threads in the warp,  and is processed in one invocation of the event handler. Warp synchronization is handled similarly.


For barrier synchronization, a barrier ID is passed with the event on a per-warp basis. Each warp involved in the barrier operation generates an event. A counter keeps track of when the last warp participating in the barrier has generated its event, and the active masks of all the participating warps are stored until this point. On detecting the final warp, the counter is reset, and the accumulated active mask information is used to correctly process the vector clocks for all active threads involved in the barrier operation. The implementations do not coalesce events for lock acquires and releases.


\paragraph*{Thread exit}

The metadata corresponding to a thread is cleared once it exits. This helps control memory overheads when a group of threads finish execution earlier and thus frees up resources. Specifically for \approach, this also helps performance as we omit pushing metadata to acquire/release queues corresponding to exited threads.

\paragraph*{Inactive threads}

A major part of the memory overhead of \Approach is from the acquire and release queues maintained
on a per-thread basis. Algorithm~\ref{algo:gwcp-acm-sync} shows that an acquire and a release
operation pushes data onto queues for every other thread. For threads that are never involved in an
acquire or a release, these queues will look identical. Thus, \approach maintains only one copy of
these queues representing the vector clock for all inactive threads and switches back to maintaining
a private copy when needed.
A thread that only accesses memory may also
be considered inactive since a memory access does not change any per-thread vector clocks when the
lockset is empty.

\paragraph*{Block-level barrier}

After a \pcode{\_\_syncthreads()}, the vector clock is identical for every thread in a
block except for the entry corresponding to a thread's local clock. This is because the final step
after a \pcode{\_\_syncthreads()} is to do an increment after taking a join with every other
thread's vector clock. This redundancy can be exploited by keeping only one copy of the vector clock for
the entire block after a \pcode{\_\_syncthreads()}, ensuring that a read of a thread's own clock
will return an appropriately incremented value. On participating in any other form of
synchronization, the implementations revert back to maintaining a private copy of the vector clock. This optimization is expected to improve memory overheads for kernels that synchronize purely with barriers, but we have not included it in the present implementation.

\smallskip
\noindent Our implementations currently use a single channel for GPU-to-CPU communication since it
is important for the CPU cores to process dependent events in the same order as the GPU execution
trace for correctness. \Approach's primary goal is to improve data race detection coverage. As a
result, we have omitted exploring additional performance optimizations such as supporting
multichannel communication between the host and the GPU, and using Unified Shared Memory to
track \gwcp relations and detect data races (\eg, \iguard~\cite{iguard-sosp-21}).
\subsection{Improving \Barracuda 
}
\label{sec:improvements}

A small contribution in this work is our extension to \barracuda to support newer (Volta onward) GPU architectures.


\paragraph*{Scoped atomics}

\Barracuda supports scoped fences and assigns a scope to each inferred lock based on the scope of
the associated fence.
Our extension supports the use of scoped atomics.
Before reporting a race, 
our implementation ensures that for two atomic accesses, the lesser of the two scopes does not
cover both accesses. An atomic access is reported as a race only if both previous and current
accesses are block-level atomics and the accesses are from different blocks.




\paragraph*{ITS}

\Barracuda models 
lockstep-based execution in pre-Volta architectures (rules \textbf{EndInsn}, \textbf{If}, and \textbf{ElseEndif} in Figure 2,~\cite{barracuda-pldi-2017}), which are broken with the introduction of ITS.
We omit these
rules and instead rely on the \code{\%laneid} PTX register~\cite{ptx-isa} and per-thread metadata for synchronization.

  \Barracuda also detects intrawarp races when warps diverge along different branches of an \pcode{if-else} statement.
  The accesses to \code{data} are racy because the interleaving order for the branches is not defined. A per-thread analysis covers such races with no extra adjustments.
  Finally, intrawarp races while executing the same instruction are also detected by our implementation.
  We iterate over all active threads in the warp at each memory access to check if two threads perform writes to the same memory location.

\swarnendu{What is ``covered''? Does it mean our implementation ``can detect''? Also, I could not make out the exact delta in the last paragraph. Please revise.}


\paragraph*{Warp-level barriers}

Volta introduces the \code{\_\_syncwarp()} synchronization primitive, a companion to ITS that forces reconvergence of a warp. We handle this primitive similar to block-level barriers by joining the per-thread vector clocks of each thread involved in the barrier.

\medskip
\noindent 
We refer to 
our extension of \barracuda as \barracudap.
We have also \emph{added} support to detect intrawarp races on the same SASS instruction to our reimplementation of \scord.


\begin{figure}[ht]
    \centering
    \begin{minipage}[t]{\linewidth}
        \begin{lstlisting}[style=CUDAStyle]
  __syncwarp();
  while (wtid == 0 &&
          atomicCAS_block(&lock, 0, 1) != 0);
  __threadfence_block();
  __syncwarp();
        \end{lstlisting}
    \end{minipage}%
    \hfill
    \caption{Common lock for entire warp.}
    \label{fig:warplock}
\end{figure}

\subsection{Vector Clock Compression}
\label{sec:compression}

\begin{figure}[t]
    \centering
    \includegraphics[scale=0.4]{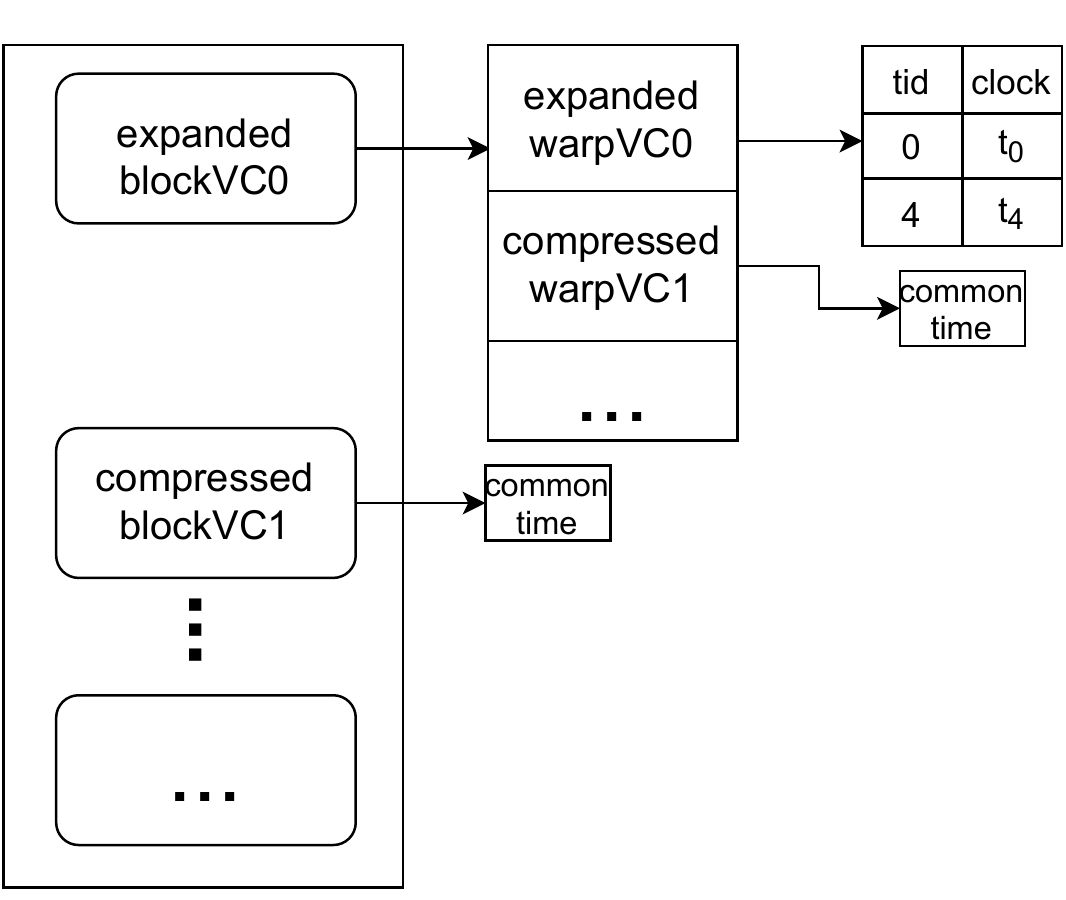}
    \caption{Generic vector clock data structure in \approach that is amenable for compression.}
    \label{fig:compression}
\end{figure}

\begin{figure}[t]
    \centering
    \includegraphics[scale=0.5]{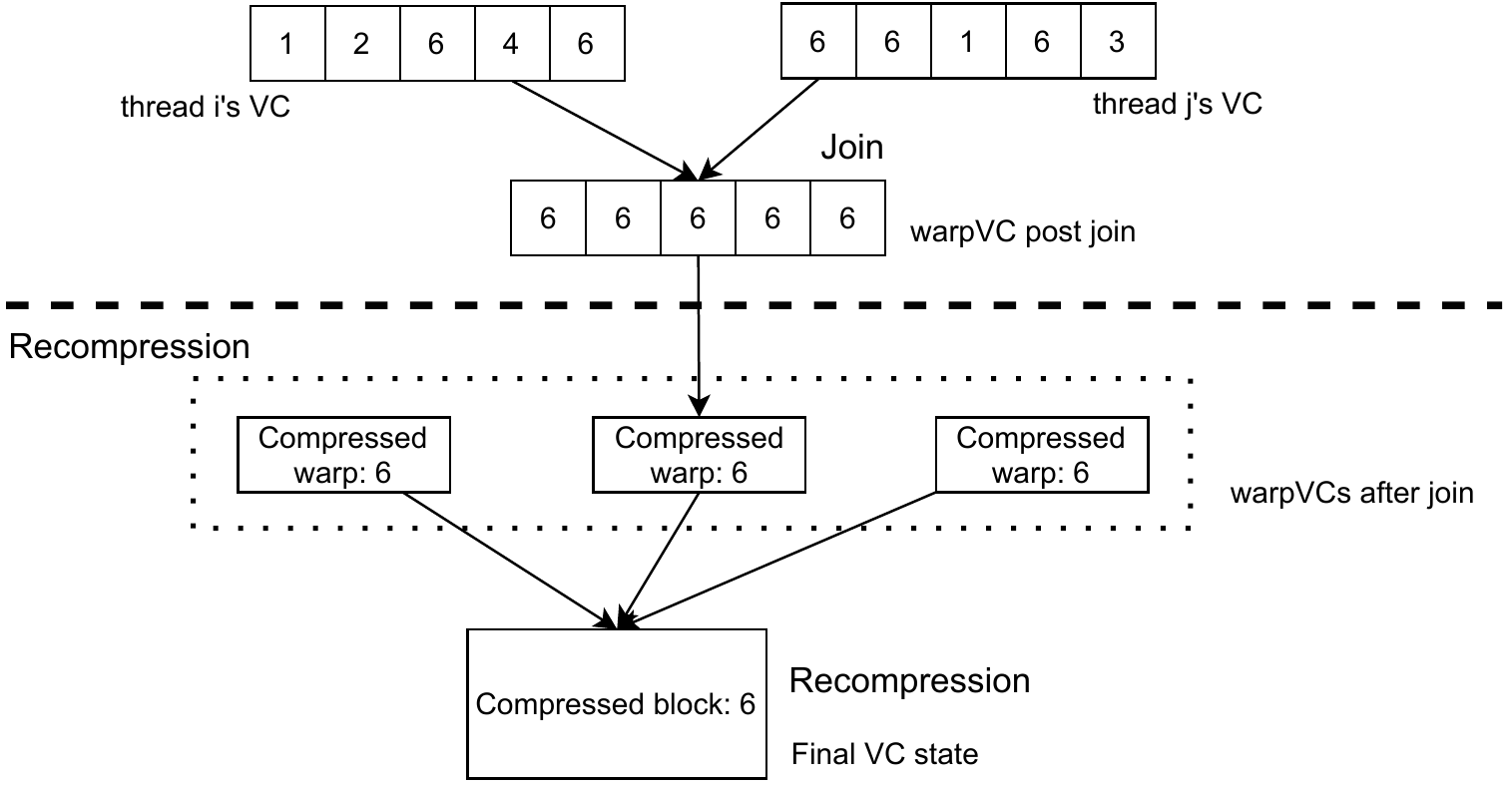}
    \caption{Re-compression operations post-join.}
    \label{fig:compressjoin}
\end{figure}

\begin{figure}[t]
    \centering
    \includegraphics[scale=0.45]{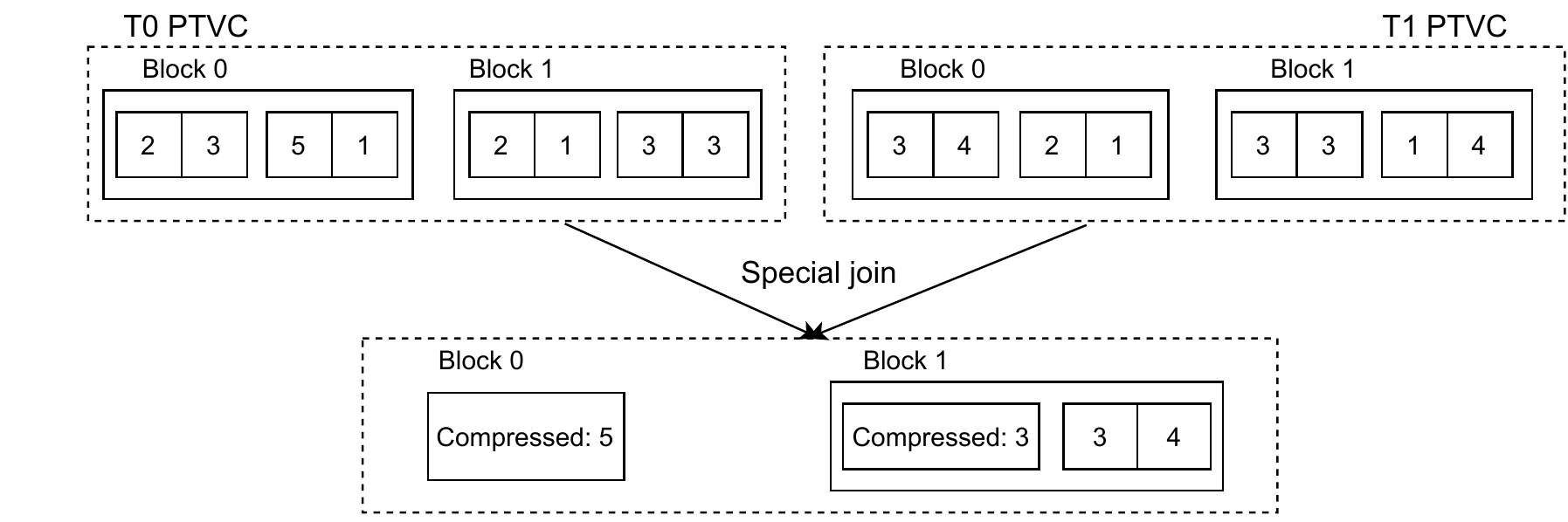}
    \caption{Illustration of a special join in case of a barrier.}
    \label{fig:Specialjoin}
\end{figure}


A \naive per-thread vector clock (PTVC) has a memory requirement of \bigO{n^{2}}, where $n$ is the
number of threads. This makes it challenging to implement vector-clock-based race detectors for GPUs
since the number of threads can be in millions.
\Barracuda
exploits redundancies in warp-level vector clocks
and stores them in a lossless compressed format whenever possible~\cite{barracuda-pldi-2017}.
Figure~\ref{fig:warplock} shows a lock acquire pattern from the \scor benchmarks
that allows a warp to acquire a single lock for the entire warp while remaining in lockstep, since warp divergence hurts performance.
The resultant vector clock will have the same logical time for each thread in a warp but will have different times across different warps.
\Barracuda's compression technique specializes to distinct vector clock states and does not exploit such redundancies.
Thus, \approach implements a more general-purpose PTVC compression logic that uses the GPU thread hierarchy to maintain compressed versions of a PTVC for any partial order.




Figure~\ref{fig:compression} shows \approach's idea for compression.
At the top level, \approach maintains an array of block-level vector clocks (denoted by \pcode{blockVC}s). Each \pcode{blockVC} can be in one of two states, \pcode{compressed} and \pcode{expanded}. In the \pcode{compressed} state, a \pcode{blockVC} maintains a common logical clock which is the same for every thread in the block. In the \pcode{expanded} state, \pcode{blockVC} is an array of warp-level vector clocks (\pcode{warpVC}s). Each \pcode{warpVC} can also be in one of two states, \pcode{compressed} and \pcode{expanded}. In the \pcode{compressed} state, a \pcode{warpVC} has a common logical time which is the same for every thread in the warp. In the \pcode{expanded} state, the \pcode{warpVC} is a map from thread ID to the (nonzero) logical clock for that thread.
\Approach 
implements two ideas to improve compression and reduce memory overheads.


\paragraph*{Re-compression attempts}

At every 
join, 
\approach checks to see if the vector clocks involved in the join can be compressed post the join.
Figure~\ref{fig:compressjoin} shows an example where post join, the warp-level vector clock (VC) is compressed. If every other VC in the block is also compressed and contains the same value, the entire block is compressed.

\paragraph*{Forced compression at barriers}

\Approach implements a novel idea to compress vector clocks at barriers. 
Instead of joining every thread at a barrier, \approach sets the vector clock entries involved in
the join to the \emph{highest} clock value among the entries. One can view this as an intrawarp join
following the standard join for the block involved in \pcode{\_\_syncthreads()}.
The update preserves ordering among threads because a barrier implies every thread involved will restart execution after reaching the barrier.
Thus, their logical times for each other can be said to be equal after the barrier completes.
For every block not involved in the 
barrier, \approach retains the normal join semantics to maintain correctness. The approach
allows for a guaranteed compression of a \pcode{blockVC} for each thread in the block. A
similar strategy can be applied for \pcode{\_\_syncwarp()}.
Figure~\ref{fig:Specialjoin} shows the sequence of compression opportunities exploited by \approach when two threads from the same block synchronize with a barrier.

\medskip
\noindent In summary, \approach's
vector clock
compression scheme is generic, naturally follows the GPU thread hierarchy, and can eliminate any redundancy found during the execution.

\section{Evaluation}
\label{sec:eval}

This section compares the data race coverage and performance of \approach with \barracudap, \scord, and \iguard.


\subsection{Experimental Setup}
\label{sec:eval:setup}

\paragraph*{Benchmarks}


We use microbenchmarks from the \scor suite\footnote{\url{https://github.com/csl-iisc/ScoR}} as
litmus tests for the correctness of our implementations and to evaluate data race coverage. Since
these benchmarks are not designed for predictable races, we have also created new microbenchmarks to
demonstrate the shortcomings of \Barracudap, \scord, and \iguard, and showcase the predictive power
of \approach. We use large applications to compare the data race coverage and the run time of
the techniques.
We use the following three applications from the \scor suite: \bench{\conv}, \bench{\red}, and \bench{\ruleone}, 
and the following seven applications from the \barracuda benchmark suite\footnote{\url{https://github.com/upenn-acg/barracuda/benchmarks}}: \bench{\hotspot}, \bench{\kmeans}, \bench{\needle}, \bench{\strmcls} (denoted by \bench{strmcls}), \bench{\pf}, \bench{shocbfs}, and \bench{\tfred} (denoted by \bench{thrfenred}). Further, we add a new benchmark, \bench{\stencil}, that showcases predictable races in large applications.
We omit benchmarks whose unmodified executables fail to run or hang with NVBit.
Finally, we omit benchmarks that take too long to run with our binary instrumentation-based implementation.




\smallskip
We will make our benchmarks and the NVBit-based implementations publicly available.



\paragraph*{Platform}






The experiments execute on an Intel Xeon Silver 4210 system with hyperthreading turned off, 128~GB DDR4 primary memory, running Ubuntu Linux 20.04.3 LTS with kernel version 5.11.0. The GPU is \NVIDIA Quadro RTX 5000 with Turing architecture and has 16~GB memory. We use
NVIDIA driver version 495.29.05 and CUDA Toolkit version 11.5. All benchmarks have been compiled for \pcode{sm\_70}.

\subsection{Data Race Coverage with Microbenchmarks}

\begin{small}
    \begin{table}
        \centering
        \begin{tabular}{lrrrrr}
                                & \textbf{Total} & \textbf{BC+} & \textbf{SRD} & \textbf{IG} & \textbf{PD} \\
            \toprule
            \pcode{norace}      & 11             & 0            & 0            & 0           & 0           \\
            \pcode{dynamic}     & 16             & 16           & 15           & 14          & 16          \\
            \pcode{predictable} & 13             & 6            & 7            & 7           & 11          \\
        \end{tabular}
        \caption{Comparison of the number of data races detected by the different dynamic analyses with our microbenchmarks.}
        \label{tab:micro-data-races}
    \end{table}
\end{small}

We classify our microbenchmarks into three categories: \pcode{norace}, \pcode{dynamic}, and \pcode{predictable}. The \pcode{dynamic} benchmarks have races in all interleavings, while the \pcode{predictable} benchmarks have their \emph{non-racy} interleavings enforced through flags. The use of flags without fences does not guarantee ordering among memory accesses, and thus the microbenchmarks contain true predictable races. Table~\ref{tab:micro-data-races} summarizes our results; BC+ denotes \barracudap, SRD denotes \scord, IG denotes \iguard, and PD denotes \approach.
The race detectors do not report any false positives.
We investigate cases where the tools miss races from the \pcode{dynamic} category. \Barracudap and \approach catch 16 out of 16 races.
There are two examples of intrawarp races:
one on the same instruction, and one on different instructions (ITS). By not modeling lockstep execution and maintaining per-thread metadata, our extended \barracudap and \approach can detect both
races. \Scord catches the first due to our extension (\Cref*{sec:improvements}), but misses the second. The \Scord implementation maintains metadata at warp granularity and performs race detection with warp identifiers, with no emphasis on individual threads. The authors recommend expanding the metadata by 5 bits to maintain thread identifiers to support ITS intrawarp races~\cite{scord-isca-2020}, but such modifications may be prohibitive on hardware platforms. While \iguard claims to handle ITS-related races, it still fails to detect the two intrawarp races
Interestingly, although \Scord catches the race in Figure~\ref{fig:iguard-fail} in every interleaving, \iguard fails to detect the race. \iguard can detect the race only if \pcode{e7} happens before \pcode{e1} in a particular run.

\begin{figure}
    \centering
    \begin{minipage}[t]{0.15\linewidth}
        \textbf{Event}
        \begin{lstlisting}[style=TraceStyle]
e1
e2
e3
e4
e5
e6
e7
        \end{lstlisting}
    \end{minipage}%
    \begin{minipage}[t]{0.35\linewidth}
        \textbf{Thread 1}
        \begin{lstlisting}[style=TraceStyle]
acq(m)
wr(x)
rel(m)




        \end{lstlisting}
    \end{minipage}%
    \begin{minipage}[t]{0.35\linewidth}
        \textbf{Thread 2}
        \begin{lstlisting}[style=TraceStyle]



acq(m)
wr(x)
rel(m)
wr(x)
        \end{lstlisting}
    \end{minipage}
    \caption{\iguard fails to find this race}
    \label{fig:iguard-fail}
\end{figure}

\paragraph*{Predictable races}

The \pcode{predictable} category has 13 races. \Barracudap can catch 6 races in programs that use just fences to order memory accesses incorrectly or have unprotected memory accesses.
It misses detecting races where the enforced interleaving introduces HB edges between the racy memory accesses.
\Scord catches such races owing to its lockset detection. However, \scord misses races that are exposed when the execution order of the critical sections is changed. \Scord also misses races between strong memory accesses separated only by a fence.
\Approach catches 11 out of the 13 races, demonstrating its effectiveness at improving race coverage for GPU programs. It catches races where unprotected accesses overlap on permuting interceding critical sections, a case which neither \Barracudap nor \Scord detect. Finally, the two races that \approach misses demonstrate \GWCP's limitations: 
\gwcp composes with HB which can hide races~\cite{smarttrack}. Furthermore, it fails if critical sections conflict early~\cite{wcp-pldi-2017}, which interestingly \scord catches due to lockset detection. The coverage of \iguard is identical to \scord in our predictable benchmarks, as is expected.


\later{\swarnendu{I wonder whether SyncP~\cite{sync-preserving-races} will find these two.}}

\subsection{Data Race Coverage with Benchmarks}

\begin{small}
    \begin{table}
        \centering
        \begin{tabular}{lrrrrr}
                              & \textbf{Total} & \textbf{BC+} & \textbf{SRD} & \textbf{IG} & \textbf{PD} \\
            \toprule
            \bench{\conv}     &                & 5            & 5            & 5           & 5           \\
            \bench{\red}      &                & 19           & 2            & 12          & 19          \\
            \bench{\ruleone}  &                & 30           & 24           & 5           & 30          \\

            \midrule

            \bench{\hotspot}  &                & 0            & 0            & 0           & 0           \\
            \bench{\kmeans}   &                & 0            & 0            & 0           & 0           \\
            \bench{\needle}   &                & 0            & 0            & 0           & 0          \\
            \bench{thrfenred} &                & 31           & 6            & 0           & 34          \\
            \bench{\pf}       &                & 2            & 2            & 0           & 5           \\
            \bench{shocbfs}   &                & 4            & 3            & 2           & 4           \\

            \midrule
            \bench{stencil}   &                & 1            & 0            & 0           & 2           \\
        \end{tabular}
        \caption{Comparison of the number of unique kinds of data races detected by the different dynamic analyses on our benchmarks.}
        \label{tab:data-races}
    \end{table}
\end{small}

\newcolumntype{H}{>{\setbox0=\hbox\bgroup}c<{\egroup}@{}}

\begin{small}
    \begin{table*}
        \centering
        \begin{tabular}{lr|rr|rrr|rr|r}
                              & \textbf{UM} & \textbf{NVB} & \textbf{BLK} & \textbf{SRD} & \textbf{BC+} & \textbf{PD} & \textbf{BC+-CP} & \textbf{PD-CP} & \textbf{IG} \\
            \toprule
            \bench{\conv}     & 0.83        & 0.83         & 1.18         & 6.53         & 10.6         & 14.7        & 11.63           & 47.69          & 7.73        \\
            \bench{\red}      & 3.18        & 3.18         & 3.42         & 18.01        & 21.93        & 42.42       & 21.24           & 47.75          & 3.59        \\
            \bench{\ruleone}  & 0.65        & 0.65         & 1.58         & 26.7         & 24.48        & 48.76       & 44.47           & 103.17         & 1.29        \\
            \midrule
            \bench{\hotspot}  & 0.35        & 0.35         & 0.62         & 3.2          & 126.84       & 232.28      & 187.41          & 517.57         & 0.67        \\
            \bench{\kmeans}   & 1.71        & 1.75         & 11.82        & 41.03        & 64.46        & 210.48      & 426.09          & 1709.39        & 2.8         \\
            \bench{\needle}   & 0.21        & 0.26         & 67.95        & 79.05        & 110.77       & 152.48      & 159.65          & 310.98         & 1.16        \\
            \bench{\pf}       & 0.23        & 0.22         & 1.55         & 3.18         & 32.63        & 95.41       & 25.27           & 52.04          & 0.54        \\
            \bench{strmcls}   & 0.28        & 0.31         & 373.74       & 393.74       & 398.93       & 420.52      & 412.91          & 425.85         & 12.17       \\
            \bench{thrfenred} & 0.22        & 0.24         & 25.99        & 39.41        & 51.37        & 88.74       & 61.73           & 122.07         & 0.99        \\
            \bench{shocbfs}   & 0.18        & 0.21         & 26.08        & 25.77        & 26.85        & 26.34       & 26.4            & 26.14          & 0.7         \\
            \midrule
            \bench{stencil}   & 0.25        & 0.25         & 0.42         & 0.42         & 0.42         & 0.42        & 0.44            & 0.49           & 0.53        \\
            \bottomrule
            \textbf{geomean}  & 1           & 1.0          & 13.8         & 35.8         & 72.3         & 97.5        & 118.5            & 195.8           & 3.6         \\
        \end{tabular}
        \caption{Comparison of the run times (in seconds) of the different dynamic analyses.}
        \label{tab:perf-overhead}
    \end{table*}
\end{small}

\newcolumntype{H}{>{\setbox0=\hbox\bgroup}c<{\egroup}@{}}

\begin{small}
  \begin{table}
    \centering
    \begin{tabular}{lrrrrr}
                        & \textbf{SRD} & \textbf{BC+} & \textbf{BC+-CP} & \textbf{PD} & \textbf{PD-CP} \\
      \toprule
      \bench{\conv}     & 494          & 1987         & 742             & 1049        & 1052           \\
      \bench{\red}      & 5016         & 6822         & 6626            & 10580       & 10218          \\
      \bench{\ruleone}  & 1065         & 2069         & 2092            & 3575        & 3570           \\
      \midrule
      \bench{\hotspot}  & 505          & 5575         & 861             & 2126        & 1338           \\
      \bench{\kmeans}   & 6580         & 6445         & 6360            & 11013       & 11004          \\
      \bench{\needle}   & 1747         & 1823         & 1829            & 3057        & 3074           \\

      \bench{\pf}       & 305          & 495          & 397             & 758         & 546            \\
      \bench{strmcls}   & 138          & 476          & 477             & 469         & 496            \\
      \bench{thrfenred} & 238          & 327          & 287             & 419         & 378            \\
      \bench{shocbfs}   & 120          & 120          & 120             & 120         & 120            \\
      \midrule
      \bench{stencil}   & 119          & 119          & 119             & 119         & 119            \\
    \end{tabular}
    \caption{Comparison of the memory overheads (in MB) of the different dynamic analyses.}
    \label{tab:memory-overhead}
  \end{table}
\end{small}

\begin{figure}[t]
  \centering
  \begin{lstlisting}[style=CUDAStyle]
  __global__ void kernel(double *dA) {
    // Initialize a constant input stencil
    const double c11, c12, c13, c21, c22, c23, c31, c32, c33;

    // Keep two values in shared memory
    __shared__ int current, lock;
    __syncthreads();

    for (int row = start; row < end; ++row) {
        // Leader thread with id 0 initializes shared memory
        if (threadIdx.x == 0) { current = 0; lock = 0; }
        __syncthreads();

        // Update current within a critical section
        while(atomicCAS(&lock, 0, 1) != 0) {}
        __threadfence();
        current = tid;
        __threadfence();
        atomicExch(&lock, 0);
        
        // Calculate the output value \sum_i \sum_j c_{ij} a_{ij}
        double value = c11 * dA[output_index - N - 1] + ...

        // Write the output back to A
        while(atomicCAS(&lock, 0, 1) != 0) {}
        __threadfence();
        dA[output_index] = value;
        __threadfence();
        atomicExch(&lock, 0);
        __syncthreads();
    }
}
        \end{lstlisting}
  \caption{Code for the \bench{\stencil} benchmark}
  \label{fig:stencil}
\end{figure}

Table~\ref{tab:data-races} summarizes the number of unique data races found by each detector.
The \bench{\stencil} benchmark contains predictable races in the interleavings we enforce. A
simplified snippet for this benchmark is shown in Figure~\ref{fig:stencil}. The benchmark performs a
stencil operation in-place. The rows of the output matrix are divided into chunks for each block to
process, with each thread processing a column. Each thread also updates a \code{current} variable
with their thread ID before starting work. In an interleaving where threads go one after the other,
read-write races on adjacent accesses are hidden by a spurious HB edge between critical sections for
updating the output and \code{current}. Furthermore, in this interleaving, all racy accesses are
separated by a fence. Therefore, while \Barracudap and \scord are unable to detect the intra-block
races, \approach catches all such races.

Across all the benchmarks, \Barracudap and \approach detect races where only a fence separates
memory accesses. The benchmark \bench{\red} contains such races, along with
write-read races on accesses separated by insufficiently-scoped fences or no fences at all. \Scord
only reports the latter two, while \Barracudap and \approach detect all three. Similar fence races exist in \bench{\ruleone}.
All four detectors detect races on insufficiently scoped atomics. \bench{\conv} contains a racy usage of a block-scoped atomic add. \bench{\ruleone} contains access patterns of the form where each thread $t$ updates indices $i_{t-1}$, $i_t$, and $i$\textsubscript{$t$+1} on an array using atomics of insufficient scope. The remaining races are
on global or shared memory;  all three detectors catch them.
\bench{thrfenred} contains global and shared memory races, along with concurrent non-atomic and atomic accesses to the same location. The races on shared memory are ITS intrawarp races, which
\scord misses.
All four tools report the remaining two race types.

\later{The race in \bench{\mm} that \Scord misses involves an invalid acquire (atomic without the following fence) with a valid release (fence followed by an atomic). Since an acquire is not inferred, \Scord resorts to its fence-based analysis, where the release fence orders memory accesses. \approach and \Barracudap do not consider this synchronization and report a race.}

\subsection{Performance Comparison}

Table~\ref{tab:perf-overhead} shows the performance of different configurations of our tools. Column \emph{UM} shows the time taken to run the unmodified program natively. \emph{NVB} shows the time taken to run the unmodified program through NVBit without instrumentation, and \emph{BLK} shows the time taken when the application is instrumented \emph{with} empty callbacks. BC+, SRD, IG, and PD denote \barracudap, \scord, \iguard, and \approach, respectively. The two columns \emph{BC+-CP} and \emph{PD-CP}, show the run times of \barracudap and \approach with our compression scheme enabled. Each value is the average of five trials.

Column BLK shows that the instrumentation with NVBit has a comparatively higher overhead for the \Barracuda benchmarks compared to the \scor benchmarks.
The overhead is especially high for \bench{strmcls} because it performs more memory accesses.
\Barracudap, \scord, and \approach incur an overhead of \bcpunmodified, \scordunmodified, and \gwcpunmodified over the native execution. The respective overheads over the BLK configuration are \bcpblank, \scordblank, and \gwcpblank, which are a more fair representation of the analyses overheads.
\Barracudap and \approach have overheads of 39.6X and 77.3X over \Scord in the worst case (for \hotspot), which primarily comes from maintaining read and write vector clocks and performing vector clock joins. \Scord is efficient since it only maintains locksets, and the number of locks used in CUDA programs are relatively less.

BC+-CP and PD-CP show the performance of \barracudap and \approach with compression.
As expected, 
compression incurs additional overhead on all benchmarks due to the extra
computation, excepting \pf. BC+-CP and PD-CP have overheads of 1.35X and 1.65X over \barracudap and
\approach respectively. The reasonable overheads of compression compared to the benefits in memory
requirements (discussed next) show that the compression schemes can be an effective choice while developing CUDA applications.


IG presents performance of the publicly available \iguard implementation. \IGUARD has very low
overhead over the unmodified application since it performs the race detection analysis in parallel
on the GPU, and does not incur the overhead of communication between the GPU and the CPU. We
emphasize that the \approach implementation focuses on race coverage; we include \iguard results for
completeness.




\paragraph*{Memory overhead}

Table~\ref{tab:memory-overhead} shows the peak memory overhead of different configurations of our tools.
The two columns, \emph{BC+-CP} and \emph{PD-CP}, show the memory overheads of \barracudap and \approach with our compression scheme enabled. Each value is the average of five trials.
We estimate the memory overhead by invoking \code{getrusage()} at the end of the execution. \scord has the lowest memory overhead compared to \barracudap and \approach.
BC+-CP has a worst-case memory overhead of less than 4X time that of \scord, with the overhead being
less than 2X of \scord in all other cases. PD has to maintain more metadata to keep track of
conflicting critical sections and thus has the highest overheads. However, PD-CP has less than 4X
the overhead of \scord in the worst case as well. The results show the generality of our compression
scheme, which significantly improves memory overheads of both \approach (up to 1.59X) and
\barracudap (up to 6.47X) analyses on the host.

\begin{figure*}[ht]
  \centering
  \begin{subfigure}{.24\textwidth}
    \centering
    \includegraphics[width=1\linewidth]{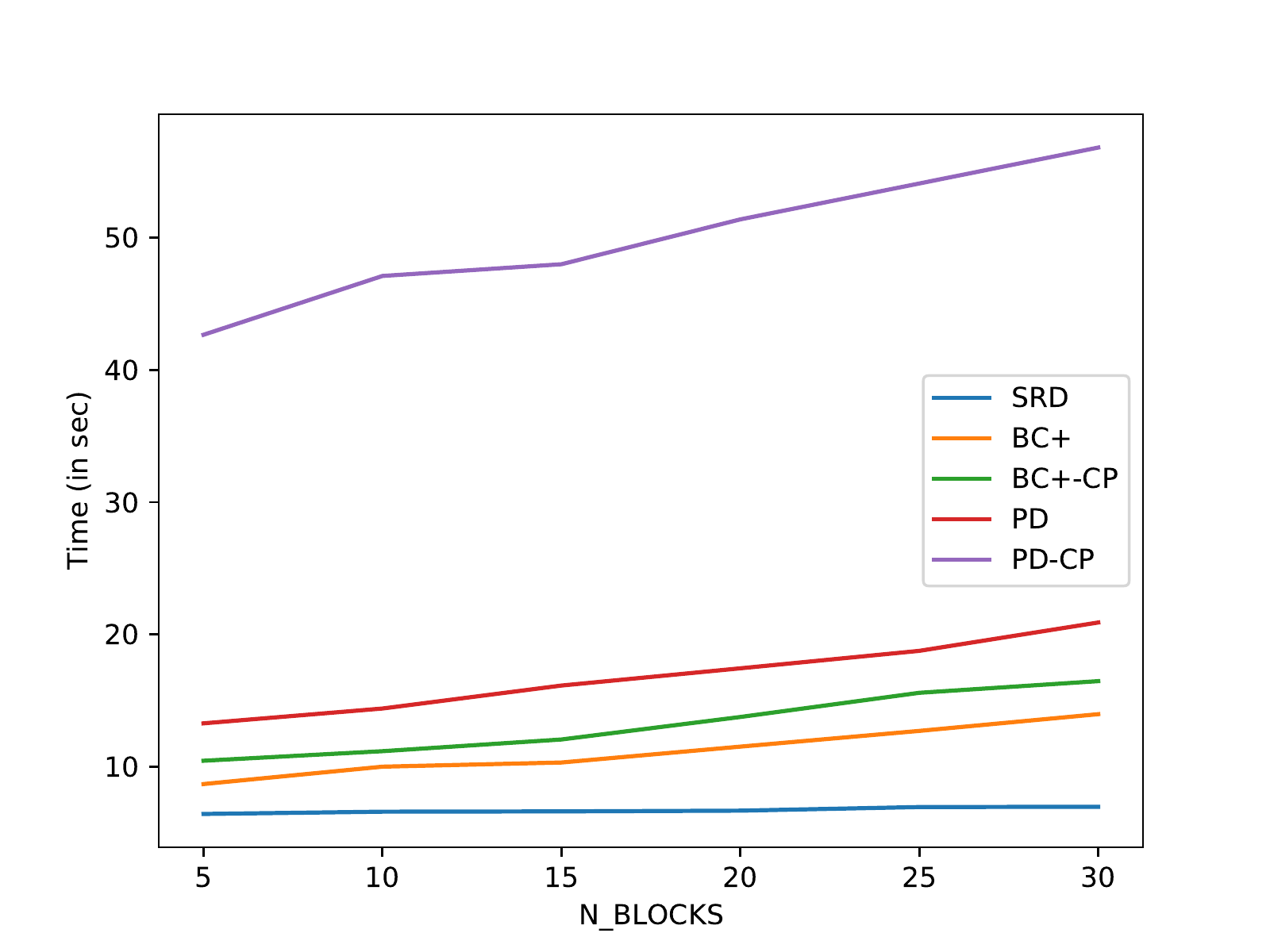}
    \caption{\conv}
    \label{fig:conv}
  \end{subfigure}%
  \begin{subfigure}{.24\textwidth}
    \centering
    \includegraphics[width=1\linewidth]{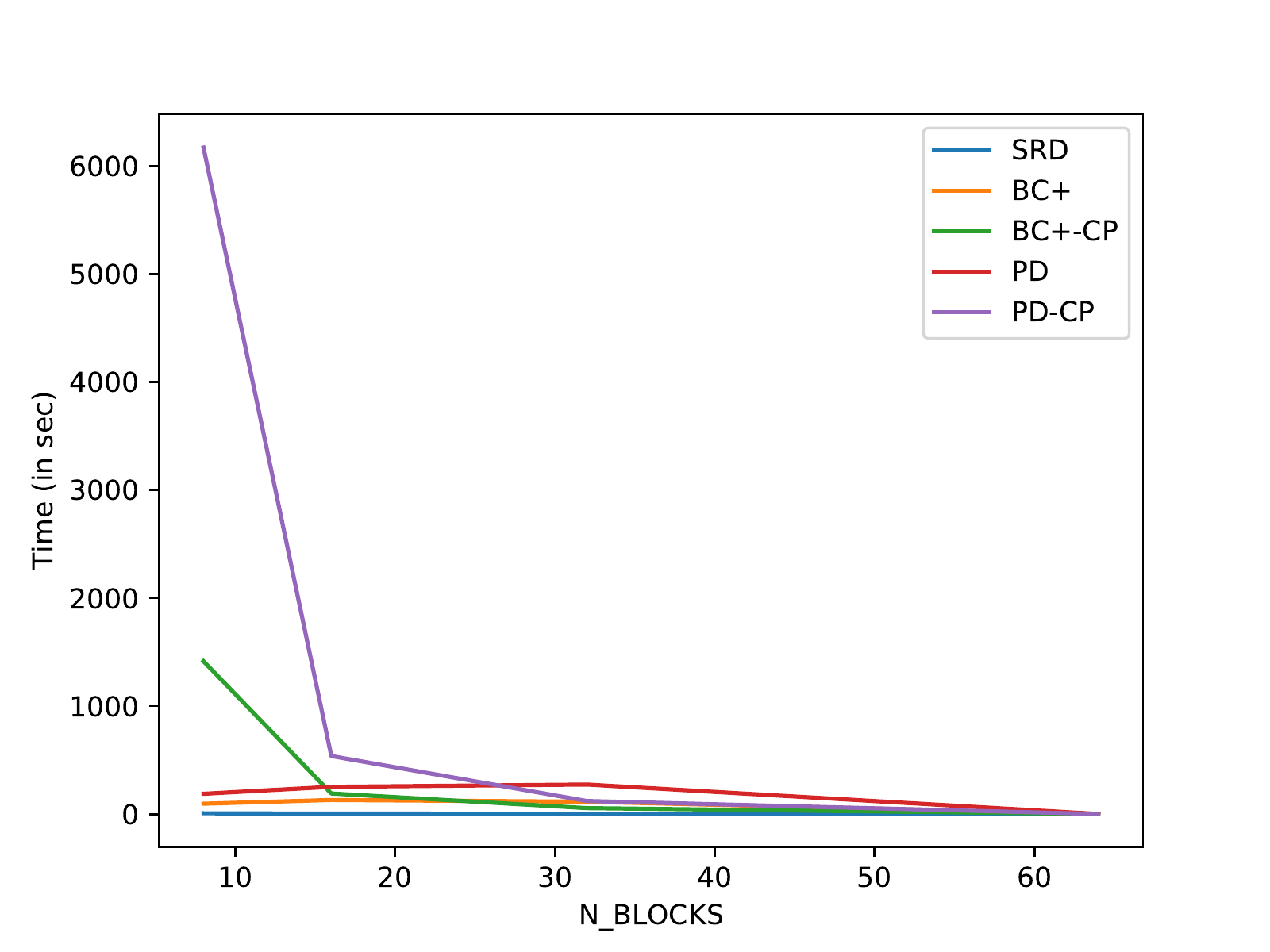}
    \caption{\hotspot}
    \label{fig:hotspot}
  \end{subfigure}
  \begin{subfigure}{.24\textwidth}
    \centering
    \includegraphics[width=1\linewidth]{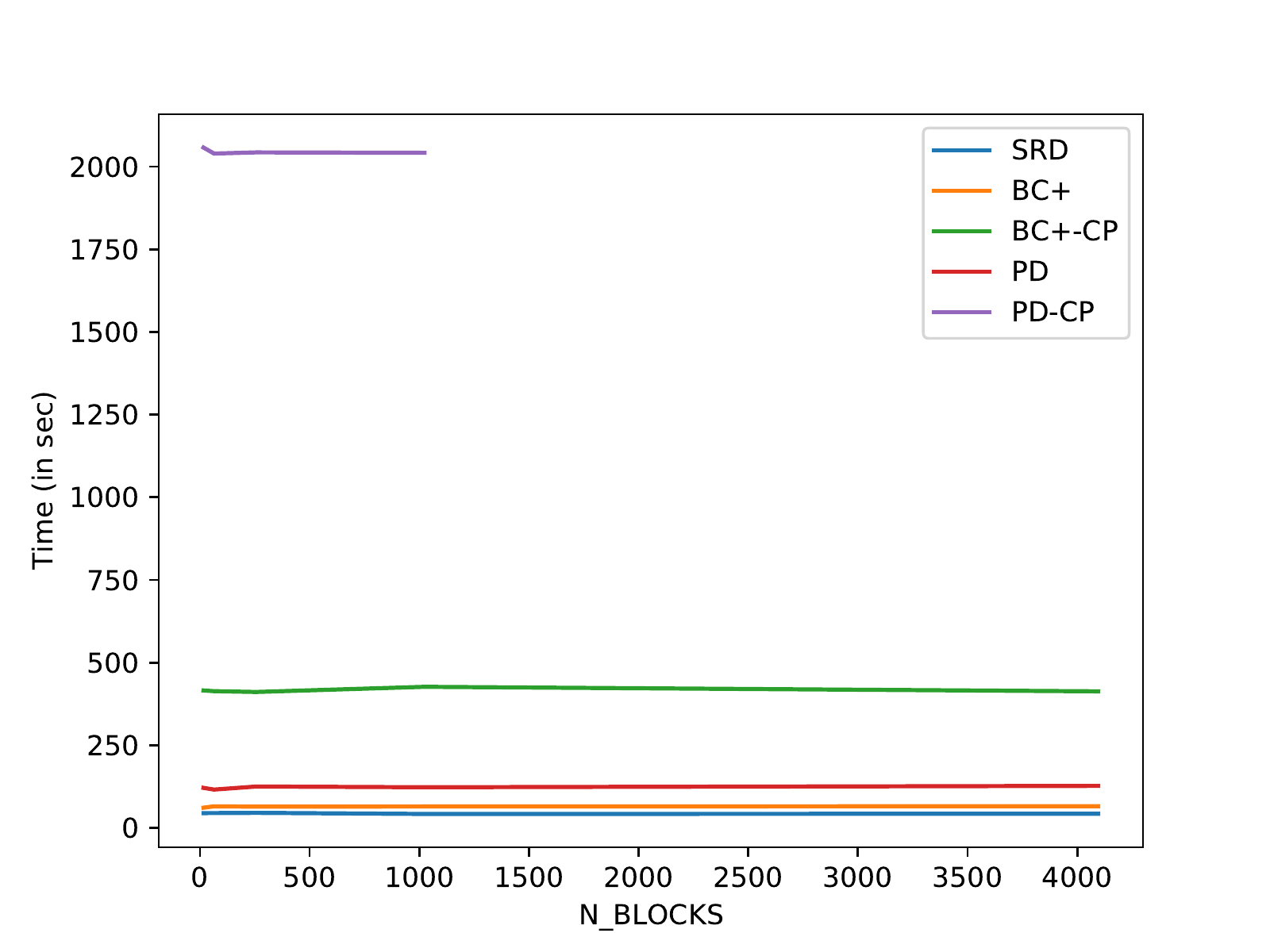}
    \caption{\kmeans}
    \label{fig:kmeans}
  \end{subfigure}
  \begin{subfigure}{.24\textwidth}
    \centering
    \includegraphics[width=1\linewidth]{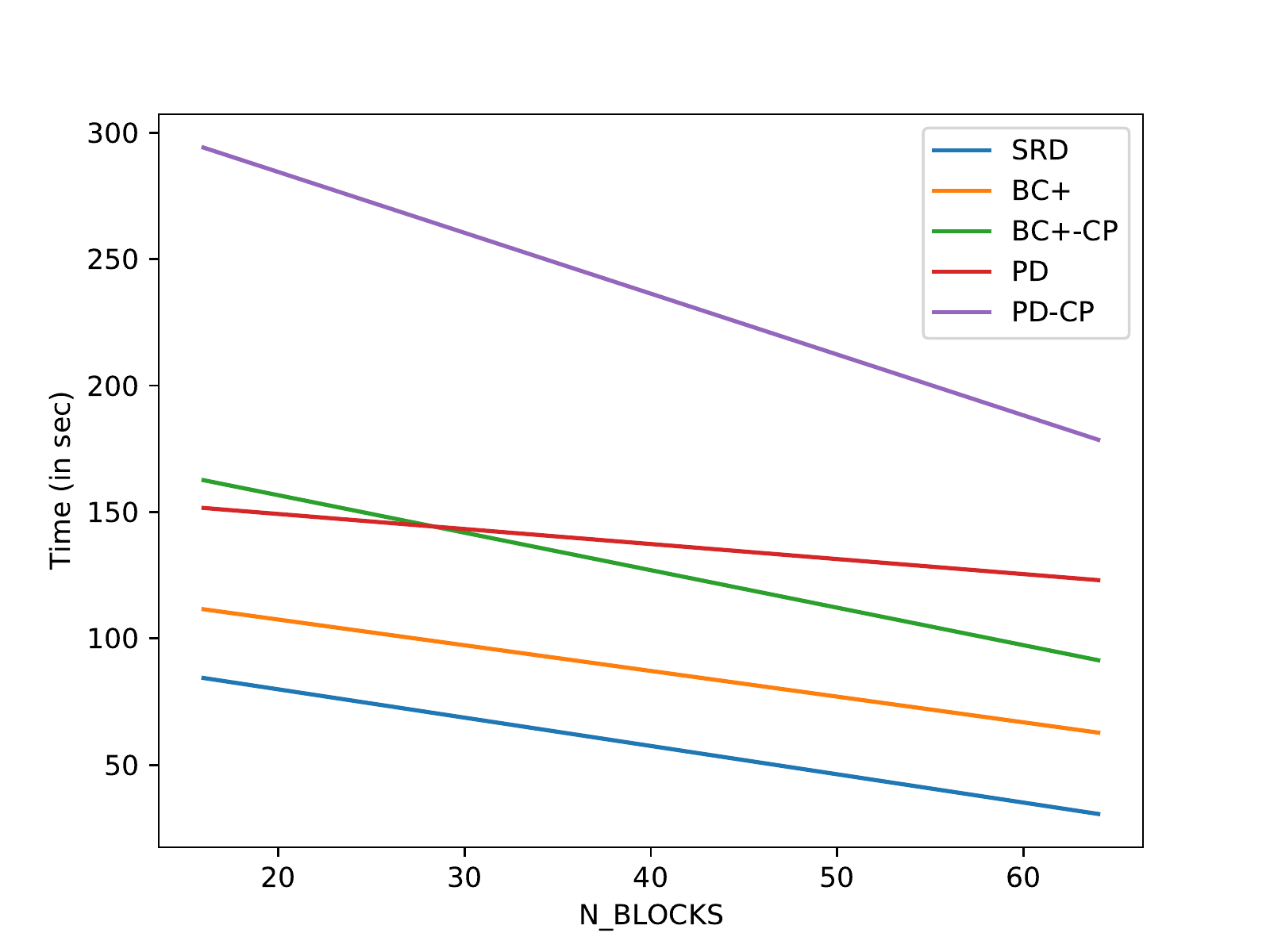}
    \caption{\needle}
    \label{fig:needle}
  \end{subfigure}\\
  \begin{subfigure}{.24\textwidth}
    \centering
    \includegraphics[width=1\linewidth]{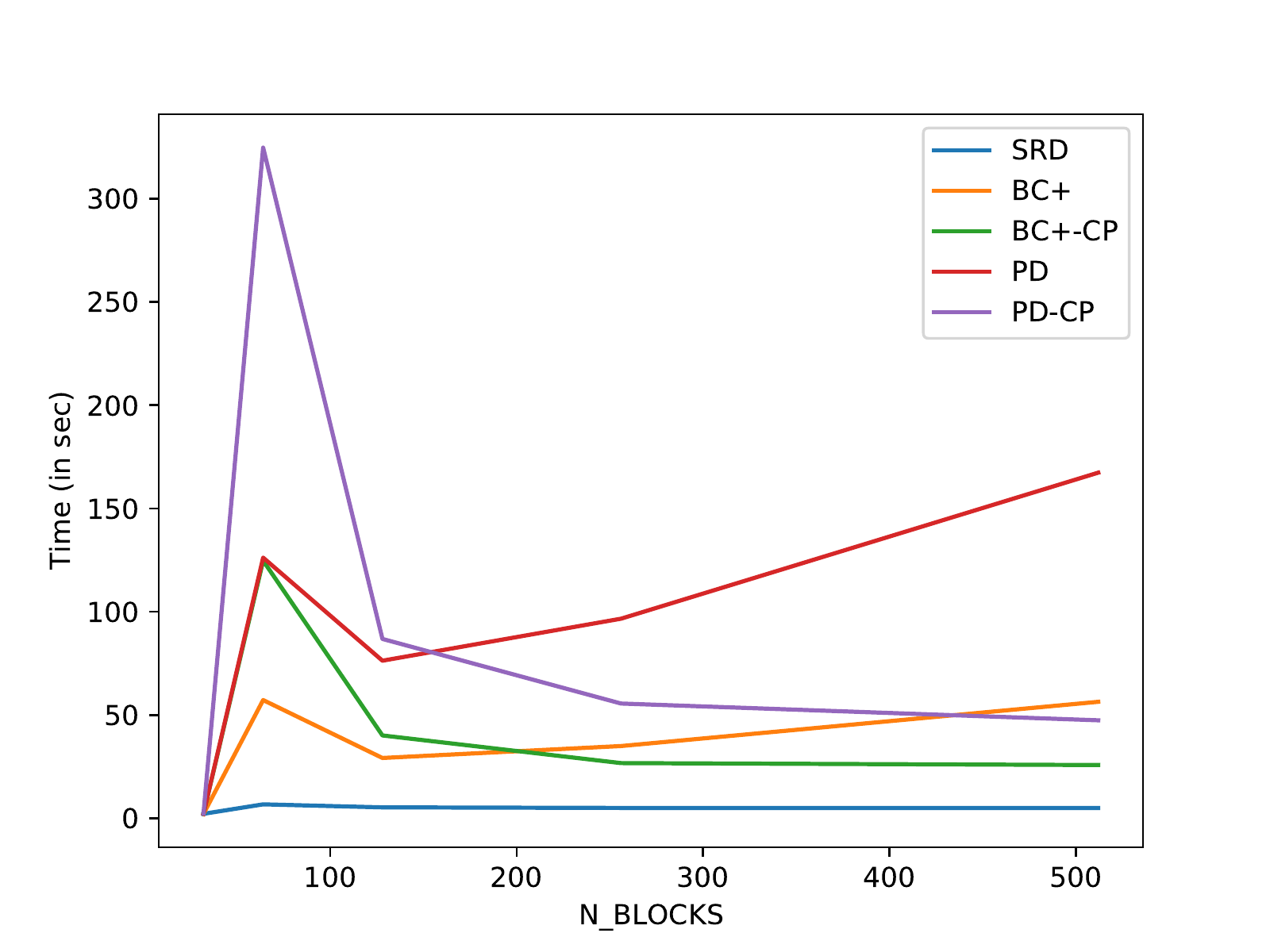}
    \caption{\pf}
    \label{fig:pathfinder}
  \end{subfigure}
  \begin{subfigure}{.24\textwidth}
    \centering
    \includegraphics[width=1\linewidth]{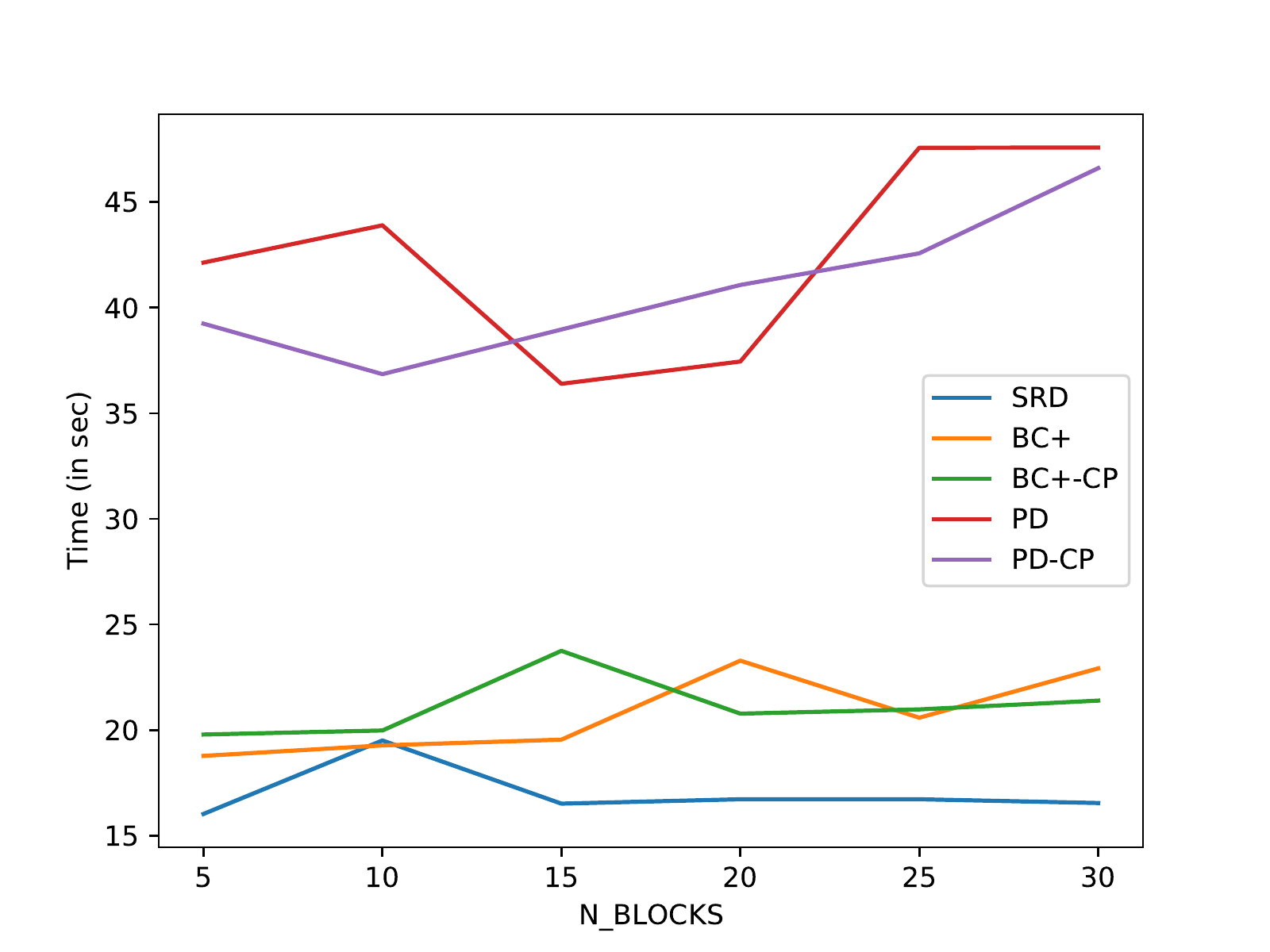}
    \caption{\red}
    \label{fig:reduction}
  \end{subfigure}
  \begin{subfigure}{.24\textwidth}
    \centering
    \includegraphics[width=1\linewidth]{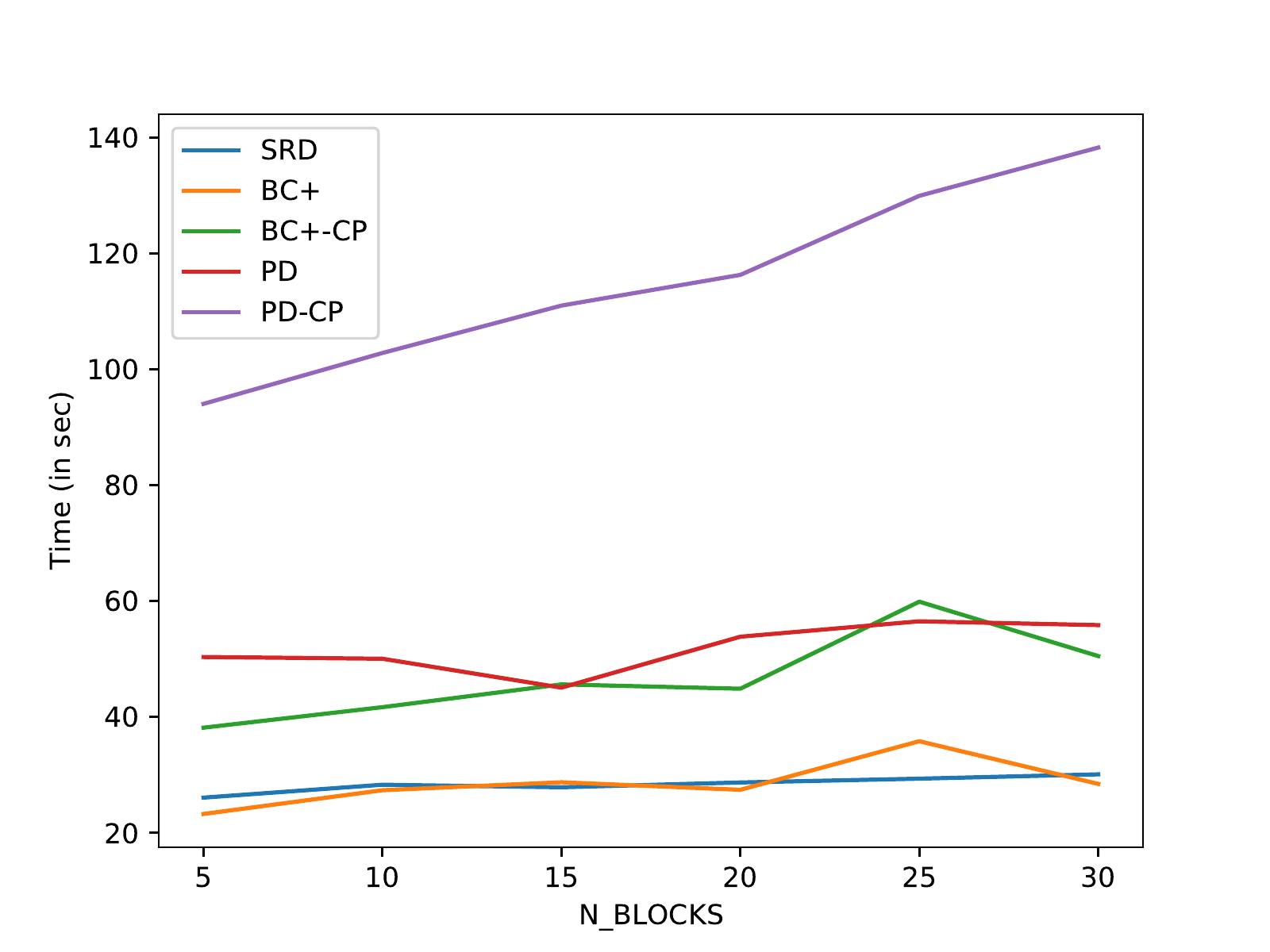}
    \caption{\ruleone}
    \label{fig:ruleone}
  \end{subfigure}
  \begin{subfigure}{.24\textwidth}
    \centering
    \includegraphics[width=1\linewidth]{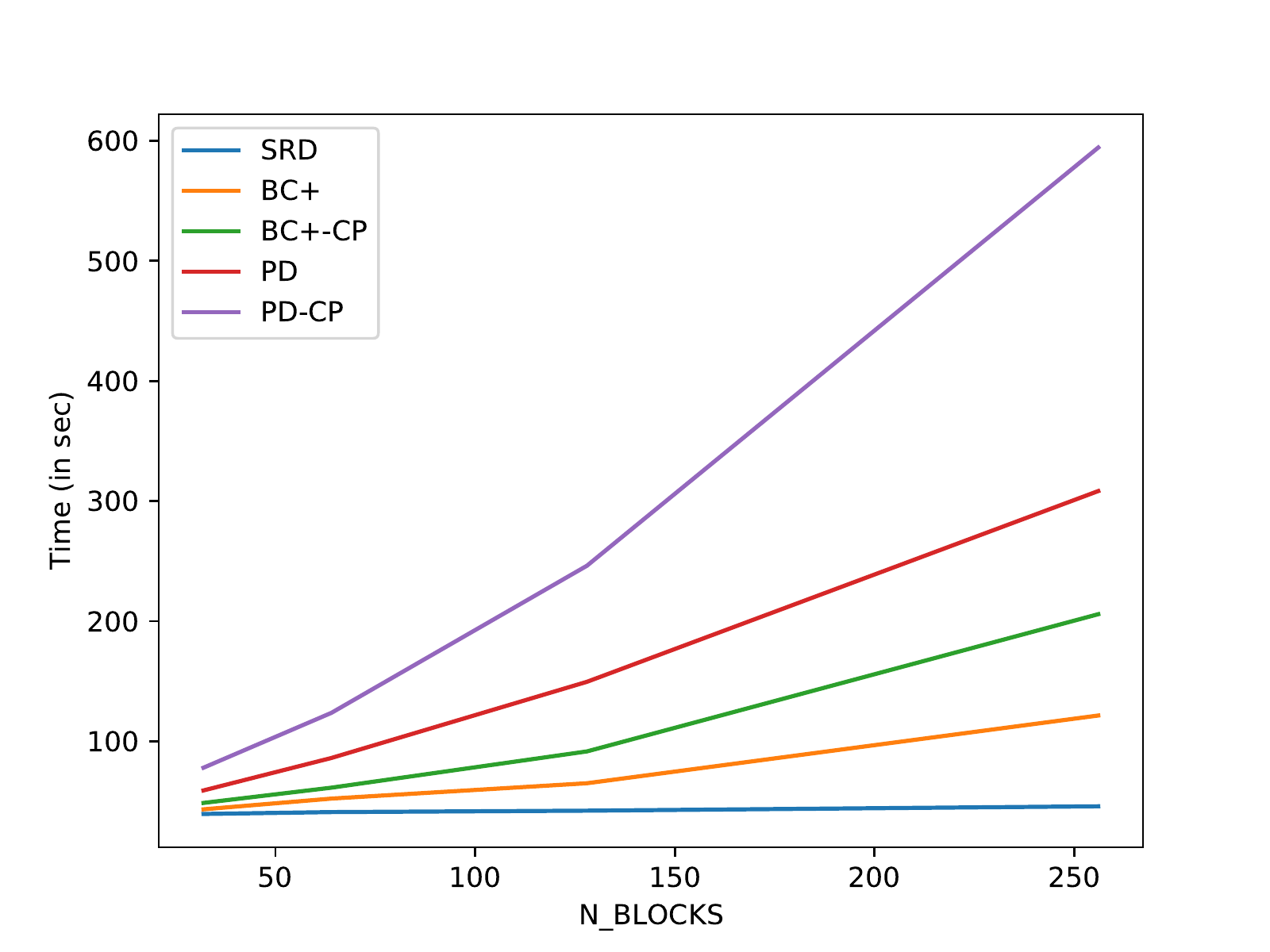}
    \caption{\bench{thrfenred}}
    \label{fig:tfred}
  \end{subfigure}\\
  \caption{Scalability results with different number of blocks.}
  \label{fig:scalability}
\end{figure*}

\paragraph*{Scalability}

Figure~\ref{fig:scalability} shows the scalability plots for those benchmarks where the number of
thread blocks can be easily configured. Note that PD-CP on \kmeans failed with 4096 blocks. \hotspot
and \needle scale well on all tools. In general, \approach has poorer scalability than \barracudap
and \scord, because of the additional computation required for predictive race detection.

\smallskip
\noindent Given that the expected use case for predictable race detectors is during application development and debugging, our experiments show that \approach provides a good data race coverage and performance tradeoff.

\section{Related Work}

In the following, we discuss other related work that has not already been discussed.

\subsection{Race Detection and Program Analyses on GPUs}

Boyer et al.~\cite{boyer-stmcs-2008} propose a dynamic analysis that instruments kernels and tracks shared memory accesses from different threads to detect data races between reads and writes (ignores write-write races).
GRace~\cite{grace-ppopp-2011} and GMRace~\cite{gmrace-tpds-2014} use static analysis to limit instrumentation so that the instrumentation overhead is reduced. These techniques separate intrawarp and interwarp race detection; the intrawarp detection logic runs after each instruction, and heavier interwarp logic runs at barriers or kernel exit.
LD~\cite{ld-taco-2017,ldetector-wodet-2014} detects data races by comparing the values of updated memory locations and avoids the overhead of synchronized metadata updates via instrumentation. LD can miss data races when the old and the new values are the same.
\emph{HaCCRgR}~\cite{haccrg-icpp-2013} uses hardware support to track cross-thread data dependences but limits tracking of concurrent readers.

Many existing race detectors detect races on shared memory accesses and ignore monitoring global
memory accesses for better performance
(\eg,~\cite{boyer-stmcs-2008,grace-ppopp-2011,gmrace-tpds-2014,cuda-racecheck}). For example, the
Racecheck tool from \NVIDIA uses dynamic binary instrumentation to detect data races on shared
memory~\cite{cuda-racecheck}. Furthermore, early work on GPU race detection assume lockstep
execution and barrier-based synchronization, and ignore synchronization with atomics or
fences~\cite{boyer-stmcs-2008,grace-ppopp-2011,gmrace-tpds-2014,ldetector-wodet-2014,ld-taco-2017}.




Program analyses of GPU kernels primarily target automated detection and fixing of synchronization and performance bugs.
{Data races are correctness problems because of incorrect synchronization, while barrier divergence and redundant barriers hurt performance.
PUG~\cite{pug-fse-2010} symbolically models barrier synchronization in kernels and uses SMT solvers to detect data races.
\emph{GPUVerify} uses SMT solvers to find data races and barrier divergence bugs~\cite{gpuverify}.
\emph{GKLEE} generates a trace of the program and uses concolic execution-based verification to identify synchronization bugs~\cite{gklee}. However, symbolic execution methods may not scale well to large input kernels and can report false alarms.
\emph{Simulee} is a dynamic analysis that generates test inputs using evolutionary programming to exercise the buggy regions of code~\cite{simulee-icse-2020}.


Compared to multithreaded shared-memory programs on CPUs, it is relatively complex to write
efficient CUDA programs and utilize the GPU memory hierarchy. Several performance profiling tools
help optimize CUDA programs~\cite{cudaadvisor,sassi-isca-2015,nvprof,ghpctoolkit}, but these
techniques do not help with concurrency correctness.


\subsection{CPU race detection}

Static data race detection techniques can potentially detect all \emph{feasible} data races across all possible executions (\ie, no false negatives), but usually do not scale to large programs and suffer from false positives~\cite{racerx,naik-static-racedet-2007,naik-static-racedet-2006,locksmith,relay-2007,racerd}.
Dynamic data race detection analyses mostly extend the popular happens-before relation to infer data races~\cite{pacer-pldi-2010,racemob,fib-oopsla-2017,google-tsan-v1,google-tsan-v2,multirace}.
Hybrid techniques integrate both HB and lockset analysis~\cite{ocallahan-hybrid-racedet-2003,racetrack}, but continue to suffer from the disadvantages of both techniques. Other techniques sacrifice soundness for performance by sampling memory accesses~\cite{datacollider,racechaser-caper-cc-2017,pacer-pldi-2010,literace,prorace} or require hardware support to speed up the race detection analysis~\cite{radish,lard,parsnip,txrace,zhou-hard}.
Data race detection analyses have also been proposed for other parallel programming models such as OpenMP~\cite{archer,romp-sc-2018,llov-taco-20,ompracer-sc-20}.

\section{Conclusion}
\label{sec:conclusion}


Designing GPU race detectors with good coverage is challenging, given the sophisticated and evolving synchronization idioms.
This work proposes (i) the \gwcp predictive partial order relation for sound and precise race detection, (ii) discusses the implementation of a tool, \approach, to track \gwcp, and (iii) discusses several optimisations to reduce memory and performance overheads in vector clock based approaches. Our evaluation shows that \approach provides a good tradeoff between the number of data races detected and the performance overhead compared to prior work.




\iftoggle{ieeeFormat}{
}{}

\begin{acks}
    We thank Aditya K. Kamath, Alvin A. George, and Arkaprava Basu for their help with the ScoRD infrastructure.

    This material is based upon work supported by IITK Initiation Grant and \grantsponsor{GS100000001}{Science and Engineering Research Board}{} under Grant \grantnum{GS100000001}{SRG/2019/000384}.
\end{acks}


\iftoggle{acmFormat}{
    \bibliographystyle{ACM-Reference-Format}
}{}
\iftoggle{asplos}{
    \bibliographystyle{plain}
}{}
\iftoggle{ieeeFormat}{
    \balance
    \bibliographystyle{IEEEtran}
}{}
\bibliography{./venue-abbrv,./references}


\begin{thebibliography}{75}


\ifx \showCODEN    \undefined \def \showCODEN     #1{\unskip}     \fi
\ifx \showDOI      \undefined \def \showDOI       #1{#1}\fi
\ifx \showISBNx    \undefined \def \showISBNx     #1{\unskip}     \fi
\ifx \showISBNxiii \undefined \def \showISBNxiii  #1{\unskip}     \fi
\ifx \showISSN     \undefined \def \showISSN      #1{\unskip}     \fi
\ifx \showLCCN     \undefined \def \showLCCN      #1{\unskip}     \fi
\ifx \shownote     \undefined \def \shownote      #1{#1}          \fi
\ifx \showarticletitle \undefined \def \showarticletitle #1{#1}   \fi
\ifx \showURL      \undefined \def \showURL       {\relax}        \fi
\providecommand\bibfield[2]{#2}
\providecommand\bibinfo[2]{#2}
\providecommand\natexlab[1]{#1}
\providecommand\showeprint[2][]{arXiv:#2}

\bibitem[\protect\citeauthoryear{{Atzeni}, {Gopalakrishnan}, {Rakamaric},
  {Ahn}, {Laguna}, {Schulz}, {Lee}, {Protze}, and {M{\"u}ller}}{{Atzeni}
  et~al\mbox{.}}{2016}]%
        {archer}
\bibfield{author}{\bibinfo{person}{S. {Atzeni}}, \bibinfo{person}{G.
  {Gopalakrishnan}}, \bibinfo{person}{Z. {Rakamaric}}, \bibinfo{person}{D.~H.
  {Ahn}}, \bibinfo{person}{I. {Laguna}}, \bibinfo{person}{M. {Schulz}},
  \bibinfo{person}{G.~L. {Lee}}, \bibinfo{person}{J. {Protze}}, {and}
  \bibinfo{person}{M.~S. {M{\"u}ller}}.} \bibinfo{year}{2016}\natexlab{}.
\newblock \showarticletitle{{ARCHER: Effectively Spotting Data Races in Large
  OpenMP Applications}}. In \bibinfo{booktitle}{\emph{IPDPS}}.
  \bibinfo{pages}{53--62}.
\newblock


\bibitem[\protect\citeauthoryear{Betts, Chong, Donaldson, Qadeer, and
  Thomson}{Betts et~al\mbox{.}}{2012}]%
        {gpuverify}
\bibfield{author}{\bibinfo{person}{Adam Betts}, \bibinfo{person}{Nathan Chong},
  \bibinfo{person}{Alastair Donaldson}, \bibinfo{person}{Shaz Qadeer}, {and}
  \bibinfo{person}{Paul Thomson}.} \bibinfo{year}{2012}\natexlab{}.
\newblock \showarticletitle{{GPUVerify: A Verifier for GPU Kernels}}. In
  \bibinfo{booktitle}{\emph{OOPSLA}}. \bibinfo{pages}{113--132}.
\newblock
\showISBNx{9781450315616}


\bibitem[\protect\citeauthoryear{Biswas, Cao, Zhang, Bond, and Wood}{Biswas
  et~al\mbox{.}}{2017}]%
        {racechaser-caper-cc-2017}
\bibfield{author}{\bibinfo{person}{Swarnendu Biswas}, \bibinfo{person}{Man
  Cao}, \bibinfo{person}{Minjia Zhang}, \bibinfo{person}{Michael~D. Bond},
  {and} \bibinfo{person}{Benjamin~P. Wood}.} \bibinfo{year}{2017}\natexlab{}.
\newblock \showarticletitle{{Lightweight Data Race Detection for Production
  Runs}}. In \bibinfo{booktitle}{\emph{CC}}. \bibinfo{pages}{11--21}.
\newblock
\showISBNx{978-1-4503-5233-8}


\bibitem[\protect\citeauthoryear{Blackshear, Gorogiannis, O'Hearn, and
  Sergey}{Blackshear et~al\mbox{.}}{2018}]%
        {racerd}
\bibfield{author}{\bibinfo{person}{Sam Blackshear}, \bibinfo{person}{Nikos
  Gorogiannis}, \bibinfo{person}{Peter~W. O'Hearn}, {and} \bibinfo{person}{Ilya
  Sergey}.} \bibinfo{year}{2018}\natexlab{}.
\newblock \showarticletitle{{RacerD: Compositional Static Race Detection}}.
\newblock \bibinfo{journal}{\emph{PACMPL}} \bibinfo{volume}{2},
  \bibinfo{number}{OOPSLA}, Article \bibinfo{articleno}{144}
  (\bibinfo{date}{Oct.} \bibinfo{year}{2018}).
\newblock


\bibitem[\protect\citeauthoryear{Bond, Coons, and McKinley}{Bond
  et~al\mbox{.}}{2010}]%
        {pacer-pldi-2010}
\bibfield{author}{\bibinfo{person}{Michael~D. Bond},
  \bibinfo{person}{Katherine~E. Coons}, {and} \bibinfo{person}{Kathryn~S.
  McKinley}.} \bibinfo{year}{2010}\natexlab{}.
\newblock \showarticletitle{{PACER: Proportional Detection of Data Races}}. In
  \bibinfo{booktitle}{\emph{PLDI}}. \bibinfo{pages}{255--268}.
\newblock
\showISBNx{978-1-4503-0019-3}


\bibitem[\protect\citeauthoryear{Bora, Das, Kukreja, Joshi, Upadrasta, and
  Rajopadhye}{Bora et~al\mbox{.}}{2020}]%
        {llov-taco-20}
\bibfield{author}{\bibinfo{person}{Utpal Bora}, \bibinfo{person}{Santanu Das},
  \bibinfo{person}{Pankaj Kukreja}, \bibinfo{person}{Saurabh Joshi},
  \bibinfo{person}{Ramakrishna Upadrasta}, {and} \bibinfo{person}{Sanjay
  Rajopadhye}.} \bibinfo{year}{2020}\natexlab{}.
\newblock \showarticletitle{{LLOV: A Fast Static Data-Race Checker for OpenMP
  Programs}}.
\newblock \bibinfo{journal}{\emph{TACO}} \bibinfo{volume}{17},
  \bibinfo{number}{4}, Article \bibinfo{articleno}{35} (\bibinfo{date}{Dec.}
  \bibinfo{year}{2020}), \bibinfo{numpages}{26}~pages.
\newblock
\showISSN{1544-3566}


\bibitem[\protect\citeauthoryear{Boyer, Skadron, and Weimer}{Boyer
  et~al\mbox{.}}{2008}]%
        {boyer-stmcs-2008}
\bibfield{author}{\bibinfo{person}{Michael Boyer}, \bibinfo{person}{Kevin
  Skadron}, {and} \bibinfo{person}{Westley Weimer}.}
  \bibinfo{year}{2008}\natexlab{}.
\newblock \showarticletitle{{Automated Dynamic Analysis of CUDA Programs}}. In
  \bibinfo{booktitle}{\emph{Workshop on Software Tools for MultiCore Systems}}.
\newblock


\bibitem[\protect\citeauthoryear{Burckhardt, Kothari, Musuvathi, and
  Nagarakatte}{Burckhardt et~al\mbox{.}}{2010}]%
        {randomized-scheduler}
\bibfield{author}{\bibinfo{person}{Sebastian Burckhardt},
  \bibinfo{person}{Pravesh Kothari}, \bibinfo{person}{Madanlal Musuvathi},
  {and} \bibinfo{person}{Santosh Nagarakatte}.}
  \bibinfo{year}{2010}\natexlab{}.
\newblock \showarticletitle{{A Randomized Scheduler with Probabilistic
  Guarantees of Finding Bugs}}. In \bibinfo{booktitle}{\emph{ASPLOS}}.
  \bibinfo{pages}{167--178}.
\newblock


\bibitem[\protect\citeauthoryear{Chabbi, Murthy, Fagan, and
  Mellor-Crummey}{Chabbi et~al\mbox{.}}{2013}]%
        {ghpctoolkit}
\bibfield{author}{\bibinfo{person}{Milind Chabbi}, \bibinfo{person}{Karthik
  Murthy}, \bibinfo{person}{Michael Fagan}, {and} \bibinfo{person}{John
  Mellor-Crummey}.} \bibinfo{year}{2013}\natexlab{}.
\newblock \showarticletitle{{Effective Sampling-Driven Performance Tools for
  GPU-Accelerated Supercomputers}}. In \bibinfo{booktitle}{\emph{SC}}. Article
  \bibinfo{articleno}{43}, \bibinfo{numpages}{12}~pages.
\newblock
\showISBNx{9781450323789}


\bibitem[\protect\citeauthoryear{Corporation}{Corporation}{2017}]%
        {volta-whitepaper}
\bibfield{author}{\bibinfo{person}{{NVIDIA} Corporation}.}
  \bibinfo{year}{2017}\natexlab{}.
\newblock \bibinfo{title}{{NVIDIA Tesla V100 GPU Architecture}}.
\newblock \bibinfo{howpublished}{Online}.
\newblock
\urldef\tempurl%
\url{https://images.nvidia.com/content/volta-architecture/pdf/volta-architecture-whitepaper.pdf}
\showURL{%
\tempurl}


\bibitem[\protect\citeauthoryear{Corporation}{Corporation}{2021}]%
        {ptx-isa}
\bibfield{author}{\bibinfo{person}{{NVIDIA} Corporation}.}
  \bibinfo{year}{2021}\natexlab{}.
\newblock \bibinfo{title}{{Parallel Thread Execution ISA}}.
\newblock \bibinfo{howpublished}{Online}.
\newblock
\urldef\tempurl%
\url{https://docs.nvidia.com/cuda/parallel-thread-execution/index.html}
\showURL{%
\tempurl}


\bibitem[\protect\citeauthoryear{Devietti, Wood, Strauss, Ceze, Grossman, and
  Qadeer}{Devietti et~al\mbox{.}}{2012}]%
        {radish}
\bibfield{author}{\bibinfo{person}{Joseph Devietti},
  \bibinfo{person}{Benjamin~P. Wood}, \bibinfo{person}{Karin Strauss},
  \bibinfo{person}{Luis Ceze}, \bibinfo{person}{Dan Grossman}, {and}
  \bibinfo{person}{Shaz Qadeer}.} \bibinfo{year}{2012}\natexlab{}.
\newblock \showarticletitle{{RADISH: Always-On Sound and Complete Race
  Detection in Software and Hardware}}. In \bibinfo{booktitle}{\emph{ISCA}}.
  \bibinfo{pages}{201--212}.
\newblock
\showISBNx{978-1-4503-1642-2}


\bibitem[\protect\citeauthoryear{Eizenberg, Peng, Pigli, Mansky, and
  Devietti}{Eizenberg et~al\mbox{.}}{2017}]%
        {barracuda-pldi-2017}
\bibfield{author}{\bibinfo{person}{Ariel Eizenberg}, \bibinfo{person}{Yuanfeng
  Peng}, \bibinfo{person}{Toma Pigli}, \bibinfo{person}{William Mansky}, {and}
  \bibinfo{person}{Joseph Devietti}.} \bibinfo{year}{2017}\natexlab{}.
\newblock \showarticletitle{{BARRACUDA: Binary-level Analysis of Runtime RAces
  in CUDA Programs}}. In \bibinfo{booktitle}{\emph{PLDI}}.
  \bibinfo{pages}{126--140}.
\newblock
\showISBNx{978-1-4503-4988-8}


\bibitem[\protect\citeauthoryear{Engler and Ashcraft}{Engler and
  Ashcraft}{2003}]%
        {racerx}
\bibfield{author}{\bibinfo{person}{Dawson Engler} {and} \bibinfo{person}{Ken
  Ashcraft}.} \bibinfo{year}{2003}\natexlab{}.
\newblock \showarticletitle{{RacerX: Effective, Static Detection of Race
  Conditions and Deadlocks}}. In \bibinfo{booktitle}{\emph{SOSP}}.
  \bibinfo{pages}{237--252}.
\newblock


\bibitem[\protect\citeauthoryear{Erickson, Musuvathi, Burckhardt, and
  Olynyk}{Erickson et~al\mbox{.}}{2010}]%
        {datacollider}
\bibfield{author}{\bibinfo{person}{John Erickson}, \bibinfo{person}{Madanlal
  Musuvathi}, \bibinfo{person}{Sebastian Burckhardt}, {and}
  \bibinfo{person}{Kirk Olynyk}.} \bibinfo{year}{2010}\natexlab{}.
\newblock \showarticletitle{{Effective Data-Race Detection for the Kernel}}. In
  \bibinfo{booktitle}{\emph{OSDI}}. \bibinfo{pages}{1--16}.
\newblock


\bibitem[\protect\citeauthoryear{Eslamimehr and Palsberg}{Eslamimehr and
  Palsberg}{2014}]%
        {racageddon}
\bibfield{author}{\bibinfo{person}{Mahdi Eslamimehr} {and}
  \bibinfo{person}{Jens Palsberg}.} \bibinfo{year}{2014}\natexlab{}.
\newblock \showarticletitle{{Race Directed Scheduling of Concurrent Programs}}.
  In \bibinfo{booktitle}{\emph{PPoPP}}. \bibinfo{pages}{301--314}.
\newblock


\bibitem[\protect\citeauthoryear{Flanagan and Freund}{Flanagan and
  Freund}{2009}]%
        {fasttrack}
\bibfield{author}{\bibinfo{person}{Cormac Flanagan} {and}
  \bibinfo{person}{Stephen~N. Freund}.} \bibinfo{year}{2009}\natexlab{}.
\newblock \showarticletitle{{FastTrack: Efficient and Precise Dynamic Race
  Detection}}. In \bibinfo{booktitle}{\emph{PLDI}}. \bibinfo{pages}{121--133}.
\newblock
\showISBNx{978-1-60558-392-1}


\bibitem[\protect\citeauthoryear{Flanagan and Freund}{Flanagan and
  Freund}{2010}]%
        {adversarial-memory}
\bibfield{author}{\bibinfo{person}{Cormac Flanagan} {and}
  \bibinfo{person}{Stephen~N. Freund}.} \bibinfo{year}{2010}\natexlab{}.
\newblock \showarticletitle{{Adversarial Memory for Detecting Destructive
  Races}}. In \bibinfo{booktitle}{\emph{PLDI}}. \bibinfo{pages}{244--254}.
\newblock
\showISBNx{978-1-4503-0019-3}


\bibitem[\protect\citeauthoryear{Gen{\c{c}}, Roemer, Xu, and Bond}{Gen{\c{c}}
  et~al\mbox{.}}{2019}]%
        {depaware-oopsla-2019}
\bibfield{author}{\bibinfo{person}{Kaan Gen{\c{c}}}, \bibinfo{person}{Jake
  Roemer}, \bibinfo{person}{Yufan Xu}, {and} \bibinfo{person}{Michael~D.
  Bond}.} \bibinfo{year}{2019}\natexlab{}.
\newblock \showarticletitle{{Dependence-Aware, Unbounded Sound Predictive Race
  Detection}}.
\newblock \bibinfo{journal}{\emph{PACMPL}} \bibinfo{volume}{3},
  \bibinfo{number}{OOPSLA}, Article \bibinfo{articleno}{179}
  (\bibinfo{date}{Oct.} \bibinfo{year}{2019}),
  \bibinfo{numpages}{179:1--179:30}~pages.
\newblock
\showISSN{2475-1421}


\bibitem[\protect\citeauthoryear{Godefroid and Nagappan}{Godefroid and
  Nagappan}{2008}]%
        {microsoft-exploratory-survey}
\bibfield{author}{\bibinfo{person}{P. Godefroid} {and} \bibinfo{person}{N.
  Nagappan}.} \bibinfo{year}{2008}\natexlab{}.
\newblock \showarticletitle{{Concurrency at Microsoft -- An Exploratory
  Survey}}. In \bibinfo{booktitle}{\emph{Workshop on Exploiting Concurrency
  Efficiently and Correctly}}.
\newblock


\bibitem[\protect\citeauthoryear{Gu and Mellor-Crummey}{Gu and
  Mellor-Crummey}{2018}]%
        {romp-sc-2018}
\bibfield{author}{\bibinfo{person}{Yizi Gu} {and} \bibinfo{person}{John
  Mellor-Crummey}.} \bibinfo{year}{2018}\natexlab{}.
\newblock \showarticletitle{{Dynamic Data Race Detection for OpenMP Programs}}.
  In \bibinfo{booktitle}{\emph{SC}}. Article \bibinfo{articleno}{61},
  \bibinfo{numpages}{12}~pages.
\newblock


\bibitem[\protect\citeauthoryear{Holey, Mekkat, and Zhai}{Holey
  et~al\mbox{.}}{2013}]%
        {haccrg-icpp-2013}
\bibfield{author}{\bibinfo{person}{Anup Holey}, \bibinfo{person}{Vineeth
  Mekkat}, {and} \bibinfo{person}{Antonia Zhai}.}
  \bibinfo{year}{2013}\natexlab{}.
\newblock \showarticletitle{{HAccRG: Hardware-Accelerated Data Race Detection
  in GPUs}}. In \bibinfo{booktitle}{\emph{ICPP}}. \bibinfo{pages}{60--69}.
\newblock


\bibitem[\protect\citeauthoryear{Huang, Meredith, and Rosu}{Huang
  et~al\mbox{.}}{2014}]%
        {rvpredict-pldi-2014}
\bibfield{author}{\bibinfo{person}{Jeff Huang}, \bibinfo{person}{Patrick~O'Neil
  Meredith}, {and} \bibinfo{person}{Grigore Rosu}.}
  \bibinfo{year}{2014}\natexlab{}.
\newblock \showarticletitle{{Maximal Sound Predictive Race Detection with
  Control Flow Abstraction}}. In \bibinfo{booktitle}{\emph{PLDI}}.
  \bibinfo{pages}{337--348}.
\newblock
\showISBNx{978-1-4503-2784-8}


\bibitem[\protect\citeauthoryear{Huang and Rajagopalan}{Huang and
  Rajagopalan}{2016}]%
        {maximal-data-race}
\bibfield{author}{\bibinfo{person}{Jeff Huang} {and} \bibinfo{person}{Arun~K.
  Rajagopalan}.} \bibinfo{year}{2016}\natexlab{}.
\newblock \showarticletitle{{Precise and Maximal Race Detection from Incomplete
  Traces}}. In \bibinfo{booktitle}{\emph{OOPSLA}}. \bibinfo{pages}{462--476}.
\newblock
\showISBNx{978-1-4503-4444-9}


\bibitem[\protect\citeauthoryear{{Jake Roemer and Michael D. Bond}}{{Jake
  Roemer and Michael D. Bond}}{2019}]%
        {raptor-arxiv}
\bibfield{author}{\bibinfo{person}{{Jake Roemer and Michael D. Bond}}.}
  \bibinfo{year}{2019}\natexlab{}.
\newblock \showarticletitle{{Online Set-Based Dynamic Analysis for Sound
  Predictive Race Detection}}.
\newblock \bibinfo{journal}{\emph{ArXiv e-prints}} (\bibinfo{date}{July}
  \bibinfo{year}{2019}).
\newblock
\showeprint[arxiv]{1907.08337}


\bibitem[\protect\citeauthoryear{Kamath and Basu}{Kamath and Basu}{2021}]%
        {iguard-sosp-21}
\bibfield{author}{\bibinfo{person}{Aditya~K. Kamath} {and}
  \bibinfo{person}{Arkaprava Basu}.} \bibinfo{year}{2021}\natexlab{}.
\newblock \showarticletitle{{iGUARD: In-GPU Advanced Race Detection}}. In
  \bibinfo{booktitle}{\emph{SOSP}}. \bibinfo{pages}{49--65}.
\newblock
\showISBNx{9781450387095}


\bibitem[\protect\citeauthoryear{Kamath, George, and Basu}{Kamath
  et~al\mbox{.}}{2020}]%
        {scord-isca-2020}
\bibfield{author}{\bibinfo{person}{Aditya~K. Kamath}, \bibinfo{person}{Alvin~A.
  George}, {and} \bibinfo{person}{Arkaprava Basu}.}
  \bibinfo{year}{2020}\natexlab{}.
\newblock \showarticletitle{{ScoRD: A Scoped Race Detector for GPUs}}. In
  \bibinfo{booktitle}{\emph{ISCA}}. \bibinfo{pages}{1036--1049}.
\newblock
\showISBNx{9781728146614}


\bibitem[\protect\citeauthoryear{Kasikci, Zamfir, and Candea}{Kasikci
  et~al\mbox{.}}{2012}]%
        {portend-asplos12}
\bibfield{author}{\bibinfo{person}{Baris Kasikci}, \bibinfo{person}{Cristian
  Zamfir}, {and} \bibinfo{person}{George Candea}.}
  \bibinfo{year}{2012}\natexlab{}.
\newblock \showarticletitle{{Data Races vs. Data Race Bugs: Telling the
  Difference with Portend}}. In \bibinfo{booktitle}{\emph{ASPLOS}}.
  \bibinfo{pages}{185--198}.
\newblock


\bibitem[\protect\citeauthoryear{Kasikci, Zamfir, and Candea}{Kasikci
  et~al\mbox{.}}{2013}]%
        {racemob}
\bibfield{author}{\bibinfo{person}{Baris Kasikci}, \bibinfo{person}{Cristian
  Zamfir}, {and} \bibinfo{person}{George Candea}.}
  \bibinfo{year}{2013}\natexlab{}.
\newblock \showarticletitle{{RaceMob: Crowdsourced Data Race Detection}}. In
  \bibinfo{booktitle}{\emph{SOSP}}. \bibinfo{pages}{406--422}.
\newblock
\showISBNx{978-1-4503-2388-8}


\bibitem[\protect\citeauthoryear{Khairy, Shen, Aamodt, and Rogers}{Khairy
  et~al\mbox{.}}{2020}]%
        {accelsim-isca-2020}
\bibfield{author}{\bibinfo{person}{Mahmoud Khairy}, \bibinfo{person}{Zhesheng
  Shen}, \bibinfo{person}{Tor~M. Aamodt}, {and} \bibinfo{person}{Timothy~G.
  Rogers}.} \bibinfo{year}{2020}\natexlab{}.
\newblock \showarticletitle{{Accel-Sim: An Extensible Simulation Framework for
  Validated GPU Modeling}}. In \bibinfo{booktitle}{\emph{ISCA}}.
  \bibinfo{pages}{473--486}.
\newblock
\showISBNx{9781728146621}


\bibitem[\protect\citeauthoryear{Kini, Mathur, and Viswanathan}{Kini
  et~al\mbox{.}}{2017}]%
        {wcp-pldi-2017}
\bibfield{author}{\bibinfo{person}{Dileep Kini}, \bibinfo{person}{Umang
  Mathur}, {and} \bibinfo{person}{Mahesh Viswanathan}.}
  \bibinfo{year}{2017}\natexlab{}.
\newblock \showarticletitle{{Dynamic Race Prediction in Linear Time}}. In
  \bibinfo{booktitle}{\emph{PLDI}}. \bibinfo{pages}{157--170}.
\newblock
\showISBNx{978-1-4503-4988-8}


\bibitem[\protect\citeauthoryear{Lamport}{Lamport}{1978}]%
        {happens-before}
\bibfield{author}{\bibinfo{person}{Leslie Lamport}.}
  \bibinfo{year}{1978}\natexlab{}.
\newblock \showarticletitle{{Time, Clocks, and the Ordering of Events in a
  Distributed System}}.
\newblock \bibinfo{journal}{\emph{CACM}} \bibinfo{volume}{21},
  \bibinfo{number}{7} (\bibinfo{year}{1978}), \bibinfo{pages}{558--565}.
\newblock
\showISSN{0001-0782}


\bibitem[\protect\citeauthoryear{Li and Gopalakrishnan}{Li and
  Gopalakrishnan}{2010}]%
        {pug-fse-2010}
\bibfield{author}{\bibinfo{person}{Guodong Li} {and} \bibinfo{person}{Ganesh
  Gopalakrishnan}.} \bibinfo{year}{2010}\natexlab{}.
\newblock \showarticletitle{{Scalable SMT-Based Verification of GPU Kernel
  Functions}}. In \bibinfo{booktitle}{\emph{FSE}}. \bibinfo{pages}{187--196}.
\newblock
\showISBNx{9781605587912}


\bibitem[\protect\citeauthoryear{Li, Li, Sawaya, Gopalakrishnan, Ghosh, and
  Rajan}{Li et~al\mbox{.}}{2012}]%
        {gklee}
\bibfield{author}{\bibinfo{person}{Guodong Li}, \bibinfo{person}{Peng Li},
  \bibinfo{person}{Geof Sawaya}, \bibinfo{person}{Ganesh Gopalakrishnan},
  \bibinfo{person}{Indradeep Ghosh}, {and} \bibinfo{person}{Sreeranga~P.
  Rajan}.} \bibinfo{year}{2012}\natexlab{}.
\newblock \showarticletitle{{GKLEE: Concolic Verification and Test Generation
  for GPUs}}. In \bibinfo{booktitle}{\emph{PPoPP}}. \bibinfo{pages}{215--224}.
\newblock
\showISBNx{9781450311601}


\bibitem[\protect\citeauthoryear{Li, Ding, Hu, and Soyata}{Li
  et~al\mbox{.}}{2014}]%
        {ldetector-wodet-2014}
\bibfield{author}{\bibinfo{person}{P. Li}, \bibinfo{person}{C. Ding},
  \bibinfo{person}{Xiaoyu Hu}, {and} \bibinfo{person}{T. Soyata}.}
  \bibinfo{year}{2014}\natexlab{}.
\newblock \showarticletitle{{LDetector: A Low Overhead Race Detector For GPU
  Programs}}. In \bibinfo{booktitle}{\emph{WoDet}}.
\newblock


\bibitem[\protect\citeauthoryear{Li, Hu, Chen, Brock, Luo, Zhang, and Ding}{Li
  et~al\mbox{.}}{2017}]%
        {ld-taco-2017}
\bibfield{author}{\bibinfo{person}{Pengcheng Li}, \bibinfo{person}{Xiaoyu Hu},
  \bibinfo{person}{Dong Chen}, \bibinfo{person}{Jacob Brock},
  \bibinfo{person}{Hao Luo}, \bibinfo{person}{Eddy~Z. Zhang}, {and}
  \bibinfo{person}{Chen Ding}.} \bibinfo{year}{2017}\natexlab{}.
\newblock \showarticletitle{{LD: Low-Overhead GPU Race Detection Without Access
  Monitoring}}.
\newblock \bibinfo{journal}{\emph{TACO}} \bibinfo{volume}{14},
  \bibinfo{number}{1}, Article \bibinfo{articleno}{9} (\bibinfo{date}{March}
  \bibinfo{year}{2017}), \bibinfo{numpages}{25}~pages.
\newblock
\showISSN{1544-3566}


\bibitem[\protect\citeauthoryear{Lu, Park, Seo, and Zhou}{Lu
  et~al\mbox{.}}{2008}]%
        {conc-bug-study-2008}
\bibfield{author}{\bibinfo{person}{Shan Lu}, \bibinfo{person}{Soyeon Park},
  \bibinfo{person}{Eunsoo Seo}, {and} \bibinfo{person}{Yuanyuan Zhou}.}
  \bibinfo{year}{2008}\natexlab{}.
\newblock \showarticletitle{{Learning from Mistakes: A Comprehensive Study on
  Real World Concurrency Bug Characteristics}}. In
  \bibinfo{booktitle}{\emph{ASPLOS}}. \bibinfo{pages}{329--339}.
\newblock
\showISBNx{978-1-59593-958-6}


\bibitem[\protect\citeauthoryear{Luo, Zou, Jin, Du, Zheng, and Shen}{Luo
  et~al\mbox{.}}{2018}]%
        {dighr}
\bibfield{author}{\bibinfo{person}{Peng Luo}, \bibinfo{person}{Deqing Zou},
  \bibinfo{person}{Hai Jin}, \bibinfo{person}{Yajuan Du}, \bibinfo{person}{Long
  Zheng}, {and} \bibinfo{person}{Jinan Shen}.} \bibinfo{year}{2018}\natexlab{}.
\newblock \showarticletitle{{DigHR: Precise Dynamic Detection of Hidden Races
  with Weak Causal Relation Analysis}}.
\newblock \bibinfo{journal}{\emph{Journal of Supercomputing}}
  \bibinfo{volume}{74}, \bibinfo{number}{6} (\bibinfo{date}{June}
  \bibinfo{year}{2018}), \bibinfo{pages}{2684--2704}.
\newblock
\showISSN{0920-8542}


\bibitem[\protect\citeauthoryear{Lustig, Sahasrabuddhe, and Giroux}{Lustig
  et~al\mbox{.}}{2019}]%
        {ptx-memory-model-nvidia}
\bibfield{author}{\bibinfo{person}{Daniel Lustig}, \bibinfo{person}{Sameer
  Sahasrabuddhe}, {and} \bibinfo{person}{Olivier Giroux}.}
  \bibinfo{year}{2019}\natexlab{}.
\newblock \showarticletitle{{A Formal Analysis of the NVIDIA PTX Memory
  Consistency Model}}. In \bibinfo{booktitle}{\emph{ASPLOS}}.
  \bibinfo{pages}{257--270}.
\newblock
\showISBNx{9781450362405}


\bibitem[\protect\citeauthoryear{Marino, Musuvathi, and Narayanasamy}{Marino
  et~al\mbox{.}}{2009}]%
        {literace}
\bibfield{author}{\bibinfo{person}{Daniel Marino}, \bibinfo{person}{Madanlal
  Musuvathi}, {and} \bibinfo{person}{Satish Narayanasamy}.}
  \bibinfo{year}{2009}\natexlab{}.
\newblock \showarticletitle{{LiteRace: Effective Sampling for Lightweight
  Data-Race Detection}}. In \bibinfo{booktitle}{\emph{PLDI}}.
  \bibinfo{pages}{134--143}.
\newblock
\showISBNx{978-1-60558-392-1}


\bibitem[\protect\citeauthoryear{Mathur, Kini, and Viswanathan}{Mathur
  et~al\mbox{.}}{2018}]%
        {shb-oopsla-2018}
\bibfield{author}{\bibinfo{person}{Umang Mathur}, \bibinfo{person}{Dileep
  Kini}, {and} \bibinfo{person}{Mahesh Viswanathan}.}
  \bibinfo{year}{2018}\natexlab{}.
\newblock \showarticletitle{{What Happens-after the First Race? Enhancing the
  Predictive Power of Happens-before Based Dynamic Race Detection}}.
\newblock \bibinfo{journal}{\emph{PACMPL}} \bibinfo{volume}{2},
  \bibinfo{number}{OOPSLA}, Article \bibinfo{articleno}{145}
  (\bibinfo{date}{Oct.} \bibinfo{year}{2018}), \bibinfo{numpages}{29}~pages.
\newblock
\showISSN{2475-1421}


\bibitem[\protect\citeauthoryear{Mathur, Pavlogiannis, and Viswanathan}{Mathur
  et~al\mbox{.}}{2021}]%
        {sync-preserving-races}
\bibfield{author}{\bibinfo{person}{Umang Mathur}, \bibinfo{person}{Andreas
  Pavlogiannis}, {and} \bibinfo{person}{Mahesh Viswanathan}.}
  \bibinfo{year}{2021}\natexlab{}.
\newblock \showarticletitle{{Optimal Prediction of Synchronization-Preserving
  Races}}.
\newblock \bibinfo{journal}{\emph{PACMPL}} \bibinfo{volume}{5},
  \bibinfo{number}{POPL}, Article \bibinfo{articleno}{36} (\bibinfo{date}{Jan.}
  \bibinfo{year}{2021}), \bibinfo{numpages}{29}~pages.
\newblock


\bibitem[\protect\citeauthoryear{Naik and Aiken}{Naik and Aiken}{2007}]%
        {naik-static-racedet-2007}
\bibfield{author}{\bibinfo{person}{Mayur Naik} {and} \bibinfo{person}{Alex
  Aiken}.} \bibinfo{year}{2007}\natexlab{}.
\newblock \showarticletitle{{Conditional Must Not Aliasing for Static Race
  Detection}}. In \bibinfo{booktitle}{\emph{POPL}}. \bibinfo{pages}{327--338}.
\newblock
\showISBNx{1-59593-575-4}


\bibitem[\protect\citeauthoryear{Naik, Aiken, and Whaley}{Naik
  et~al\mbox{.}}{2006}]%
        {naik-static-racedet-2006}
\bibfield{author}{\bibinfo{person}{Mayur Naik}, \bibinfo{person}{Alex Aiken},
  {and} \bibinfo{person}{John Whaley}.} \bibinfo{year}{2006}\natexlab{}.
\newblock \showarticletitle{{Effective Static Race Detection for Java}}. In
  \bibinfo{booktitle}{\emph{PLDI}}. \bibinfo{pages}{308--319}.
\newblock
\showISBNx{1-59593-320-4}


\bibitem[\protect\citeauthoryear{NVIDIA}{NVIDIA}{2021a}]%
        {cuda-programming-guide}
\bibfield{author}{\bibinfo{person}{NVIDIA}.} \bibinfo{year}{2021}\natexlab{a}.
\newblock \bibinfo{title}{{CUDA {C++} Programming Guide}}.
\newblock \bibinfo{howpublished}{Online}.
\newblock
\urldef\tempurl%
\url{https://docs.nvidia.com/cuda/cuda-c-programming-guide/index.html}
\showURL{%
\tempurl}


\bibitem[\protect\citeauthoryear{NVIDIA}{NVIDIA}{2021b}]%
        {cuda-racecheck}
\bibfield{author}{\bibinfo{person}{NVIDIA}.} \bibinfo{year}{2021}\natexlab{b}.
\newblock \bibinfo{title}{{Racecheck Tool}}.
\newblock \bibinfo{howpublished}{Online}.
\newblock
\urldef\tempurl%
\url{https://docs.nvidia.com/cuda/cuda-memcheck/index.html#racecheck-tool}
\showURL{%
\tempurl}


\bibitem[\protect\citeauthoryear{{NVIDIA Developer}}{{NVIDIA
  Developer}}{2021}]%
        {nvprof}
\bibfield{author}{\bibinfo{person}{{NVIDIA Developer}}.}
  \bibinfo{year}{2021}\natexlab{}.
\newblock \bibinfo{title}{{Profiler: CUDA Toolkit Documentation}}.
\newblock \bibinfo{howpublished}{Online}.
\newblock
\urldef\tempurl%
\url{https://docs.nvidia.com/cuda/profiler-users-guide/}
\showURL{%
\tempurl}


\bibitem[\protect\citeauthoryear{O'Callahan and Choi}{O'Callahan and
  Choi}{2003}]%
        {ocallahan-hybrid-racedet-2003}
\bibfield{author}{\bibinfo{person}{Robert O'Callahan} {and}
  \bibinfo{person}{Jong-Deok Choi}.} \bibinfo{year}{2003}\natexlab{}.
\newblock \showarticletitle{{Hybrid Dynamic Data Race Detection}}. In
  \bibinfo{booktitle}{\emph{PPoPP}}. \bibinfo{pages}{167--178}.
\newblock


\bibitem[\protect\citeauthoryear{Pavlogiannis}{Pavlogiannis}{2019}]%
        {pavlogiannis-2019}
\bibfield{author}{\bibinfo{person}{Andreas Pavlogiannis}.}
  \bibinfo{year}{2019}\natexlab{}.
\newblock \showarticletitle{{Fast, Sound, and Effectively Complete Dynamic Race
  Prediction}}.
\newblock \bibinfo{journal}{\emph{PACMPL}} \bibinfo{volume}{4},
  \bibinfo{number}{POPL}, Article \bibinfo{articleno}{17}
  (\bibinfo{year}{2019}).
\newblock


\bibitem[\protect\citeauthoryear{Peng, Grover, and Devietti}{Peng
  et~al\mbox{.}}{2018}]%
        {curd-pldi-2018}
\bibfield{author}{\bibinfo{person}{Yuanfeng Peng}, \bibinfo{person}{Vinod
  Grover}, {and} \bibinfo{person}{Joseph Devietti}.}
  \bibinfo{year}{2018}\natexlab{}.
\newblock \showarticletitle{{CURD: A Dynamic CUDA Race Detector}}. In
  \bibinfo{booktitle}{\emph{PLDI}}. \bibinfo{pages}{390--403}.
\newblock
\showISBNx{978-1-4503-5698-5}


\bibitem[\protect\citeauthoryear{Peng, Wood, and Devietti}{Peng
  et~al\mbox{.}}{2017}]%
        {parsnip}
\bibfield{author}{\bibinfo{person}{Yuanfeng Peng}, \bibinfo{person}{Benjamin~P.
  Wood}, {and} \bibinfo{person}{Joseph Devietti}.}
  \bibinfo{year}{2017}\natexlab{}.
\newblock \showarticletitle{{PARSNIP: Performant Architecture for Race Safety
  with No Impact on Precision}}. In \bibinfo{booktitle}{\emph{MICRO}}.
  \bibinfo{pages}{490--502}.
\newblock
\showISBNx{978-1-4503-4952-9}


\bibitem[\protect\citeauthoryear{Pozniansky and Schuster}{Pozniansky and
  Schuster}{2007}]%
        {multirace}
\bibfield{author}{\bibinfo{person}{Eli Pozniansky} {and} \bibinfo{person}{Assaf
  Schuster}.} \bibinfo{year}{2007}\natexlab{}.
\newblock \showarticletitle{{MultiRace: Efficient On-the-Fly Data Race
  Detection in Multithreaded C++ Programs}}.
\newblock \bibinfo{journal}{\emph{CCPE}} \bibinfo{volume}{19},
  \bibinfo{number}{3} (\bibinfo{year}{2007}), \bibinfo{pages}{327--340}.
\newblock
\showISSN{1532-0626}


\bibitem[\protect\citeauthoryear{Pratikakis, Foster, and Hicks}{Pratikakis
  et~al\mbox{.}}{2006}]%
        {locksmith}
\bibfield{author}{\bibinfo{person}{Polyvios Pratikakis},
  \bibinfo{person}{Jeffrey~S. Foster}, {and} \bibinfo{person}{Michael Hicks}.}
  \bibinfo{year}{2006}\natexlab{}.
\newblock \showarticletitle{{LOCKSMITH: Context-Sensitive Correlation Analysis
  for Race Detection}}. In \bibinfo{booktitle}{\emph{PLDI}}.
  \bibinfo{pages}{320--331}.
\newblock


\bibitem[\protect\citeauthoryear{Roemer, Gen\c{c}, and Bond}{Roemer
  et~al\mbox{.}}{2018}]%
        {vindicator-pldi-2018}
\bibfield{author}{\bibinfo{person}{Jake Roemer}, \bibinfo{person}{Kaan
  Gen\c{c}}, {and} \bibinfo{person}{Michael~D. Bond}.}
  \bibinfo{year}{2018}\natexlab{}.
\newblock \showarticletitle{{High-coverage, Unbounded Sound Predictive Race
  Detection}}. In \bibinfo{booktitle}{\emph{PLDI}}. \bibinfo{pages}{374--389}.
\newblock
\showISBNx{978-1-4503-5698-5}


\bibitem[\protect\citeauthoryear{Roemer, Gen\c{c}, and Bond}{Roemer
  et~al\mbox{.}}{2020}]%
        {smarttrack}
\bibfield{author}{\bibinfo{person}{Jake Roemer}, \bibinfo{person}{Kaan
  Gen\c{c}}, {and} \bibinfo{person}{Michael~D. Bond}.}
  \bibinfo{year}{2020}\natexlab{}.
\newblock \showarticletitle{{SmartTrack: Efficient Predictive Race Detection}}.
  In \bibinfo{booktitle}{\emph{PLDI}}. \bibinfo{pages}{747--762}.
\newblock
\showISBNx{9781450376136}


\bibitem[\protect\citeauthoryear{Savage, Burrows, Nelson, Sobalvarro, and
  Anderson}{Savage et~al\mbox{.}}{1997}]%
        {eraser}
\bibfield{author}{\bibinfo{person}{Stefan Savage}, \bibinfo{person}{Michael
  Burrows}, \bibinfo{person}{Greg Nelson}, \bibinfo{person}{Patrick
  Sobalvarro}, {and} \bibinfo{person}{Thomas Anderson}.}
  \bibinfo{year}{1997}\natexlab{}.
\newblock \showarticletitle{{Eraser: A Dynamic Data Race Detector for
  Multi-threaded Programs}}. In \bibinfo{booktitle}{\emph{SOSP}}.
  \bibinfo{pages}{27--37}.
\newblock
\showISBNx{0-89791-916-5}


\bibitem[\protect\citeauthoryear{Sen}{Sen}{2008}]%
        {racefuzzer}
\bibfield{author}{\bibinfo{person}{Koushik Sen}.}
  \bibinfo{year}{2008}\natexlab{}.
\newblock \showarticletitle{{Race Directed Random Testing of Concurrent
  Programs}}. In \bibinfo{booktitle}{\emph{PLDI}}. \bibinfo{pages}{11--21}.
\newblock


\bibitem[\protect\citeauthoryear{Serebryany and Iskhodzhanov}{Serebryany and
  Iskhodzhanov}{2009}]%
        {google-tsan-v1}
\bibfield{author}{\bibinfo{person}{Konstantin Serebryany} {and}
  \bibinfo{person}{Timur Iskhodzhanov}.} \bibinfo{year}{2009}\natexlab{}.
\newblock \showarticletitle{{ThreadSanitizer -- data race detection in
  practice}}. In \bibinfo{booktitle}{\emph{WBIA}}. \bibinfo{pages}{62--71}.
\newblock


\bibitem[\protect\citeauthoryear{Serebryany, Potapenko, Iskhodzhanov, and
  Vyukov}{Serebryany et~al\mbox{.}}{2012}]%
        {google-tsan-v2}
\bibfield{author}{\bibinfo{person}{Konstantin Serebryany},
  \bibinfo{person}{Alexander Potapenko}, \bibinfo{person}{Timur Iskhodzhanov},
  {and} \bibinfo{person}{Dmitriy Vyukov}.} \bibinfo{year}{2012}\natexlab{}.
\newblock \showarticletitle{{Dynamic Race Detection with LLVM Compiler}}. In
  \bibinfo{booktitle}{\emph{RV}}. \bibinfo{pages}{110--114}.
\newblock
\showISBNx{978-3-642-29859-2}


\bibitem[\protect\citeauthoryear{Shen, Song, Li, and Liu}{Shen
  et~al\mbox{.}}{2018}]%
        {cudaadvisor}
\bibfield{author}{\bibinfo{person}{Du Shen}, \bibinfo{person}{Shuaiwen~Leon
  Song}, \bibinfo{person}{Ang Li}, {and} \bibinfo{person}{Xu Liu}.}
  \bibinfo{year}{2018}\natexlab{}.
\newblock \showarticletitle{{CUDAAdvisor: LLVM-Based Runtime Profiling for
  Modern GPUs}}. In \bibinfo{booktitle}{\emph{CGO}}. \bibinfo{pages}{214--227}.
\newblock
\showISBNx{9781450356176}


\bibitem[\protect\citeauthoryear{Smaragdakis, Evans, Sadowski, Yi, and
  Flanagan}{Smaragdakis et~al\mbox{.}}{2012}]%
        {cp-popl-2012}
\bibfield{author}{\bibinfo{person}{Yannis Smaragdakis}, \bibinfo{person}{Jacob
  Evans}, \bibinfo{person}{Caitlin Sadowski}, \bibinfo{person}{Jaeheon Yi},
  {and} \bibinfo{person}{Cormac Flanagan}.} \bibinfo{year}{2012}\natexlab{}.
\newblock \showarticletitle{{Sound Predictive Race Detection in Polynomial
  Time}}. In \bibinfo{booktitle}{\emph{POPL}}. \bibinfo{pages}{387--400}.
\newblock
\showISBNx{978-1-4503-1083-3}


\bibitem[\protect\citeauthoryear{Stephenson, Sastry~Hari, Lee, Ebrahimi,
  Johnson, Nellans, O'Connor, and Keckler}{Stephenson et~al\mbox{.}}{2015}]%
        {sassi-isca-2015}
\bibfield{author}{\bibinfo{person}{Mark Stephenson},
  \bibinfo{person}{Siva~Kumar Sastry~Hari}, \bibinfo{person}{Yunsup Lee},
  \bibinfo{person}{Eiman Ebrahimi}, \bibinfo{person}{Daniel~R. Johnson},
  \bibinfo{person}{David Nellans}, \bibinfo{person}{Mike O'Connor}, {and}
  \bibinfo{person}{Stephen~W. Keckler}.} \bibinfo{year}{2015}\natexlab{}.
\newblock \showarticletitle{{Flexible Software Profiling of GPU
  Architectures}}. In \bibinfo{booktitle}{\emph{ISCA}}.
  \bibinfo{pages}{185--197}.
\newblock
\showISBNx{9781450334020}


\bibitem[\protect\citeauthoryear{Stone, Gohara, and Shi}{Stone
  et~al\mbox{.}}{2010}]%
        {opencl}
\bibfield{author}{\bibinfo{person}{J. Stone}, \bibinfo{person}{D. Gohara},
  {and} \bibinfo{person}{Guochun Shi}.} \bibinfo{year}{2010}\natexlab{}.
\newblock \showarticletitle{{OpenCL: A Parallel Programming Standard for
  Heterogeneous Computing Systems}}.
\newblock \bibinfo{journal}{\emph{Computing in Science \& Engineering}}
  \bibinfo{volume}{12} (\bibinfo{date}{May} \bibinfo{year}{2010}),
  \bibinfo{pages}{66--72}.
\newblock


\bibitem[\protect\citeauthoryear{Swain, Li, Liu, Laguna, Georgakoudis, and
  Huang}{Swain et~al\mbox{.}}{2020}]%
        {ompracer-sc-20}
\bibfield{author}{\bibinfo{person}{Bradley Swain}, \bibinfo{person}{Yanze Li},
  \bibinfo{person}{Peiming Liu}, \bibinfo{person}{Ignacio Laguna},
  \bibinfo{person}{Giorgis Georgakoudis}, {and} \bibinfo{person}{Jeff Huang}.}
  \bibinfo{year}{2020}\natexlab{}.
\newblock \showarticletitle{{OMPRacer: A Scalable and Precise Static Race
  Detector for OpenMP Programs}}. In \bibinfo{booktitle}{\emph{SC}}. Article
  \bibinfo{articleno}{54}, \bibinfo{numpages}{14}~pages.
\newblock
\showISBNx{9781728199986}


\bibitem[\protect\citeauthoryear{Villa, Stephenson, Nellans, and Keckler}{Villa
  et~al\mbox{.}}{2019}]%
        {nvbit-micro-2019}
\bibfield{author}{\bibinfo{person}{Oreste Villa}, \bibinfo{person}{Mark
  Stephenson}, \bibinfo{person}{David Nellans}, {and}
  \bibinfo{person}{Stephen~W. Keckler}.} \bibinfo{year}{2019}\natexlab{}.
\newblock \showarticletitle{{NVBit: A Dynamic Binary Instrumentation Framework
  for NVIDIA GPUs}}. In \bibinfo{booktitle}{\emph{MICRO}}.
  \bibinfo{pages}{372--383}.
\newblock
\showISBNx{9781450369381}


\bibitem[\protect\citeauthoryear{Voung, Jhala, and Lerner}{Voung
  et~al\mbox{.}}{2007}]%
        {relay-2007}
\bibfield{author}{\bibinfo{person}{Jan~Wen Voung}, \bibinfo{person}{Ranjit
  Jhala}, {and} \bibinfo{person}{Sorin Lerner}.}
  \bibinfo{year}{2007}\natexlab{}.
\newblock \showarticletitle{{RELAY: Static Race Detection on Millions of Lines
  of Code}}. In \bibinfo{booktitle}{\emph{ESEC/FSE}}.
  \bibinfo{pages}{205--214}.
\newblock


\bibitem[\protect\citeauthoryear{Wood, Cao, Bond, and Grossman}{Wood
  et~al\mbox{.}}{2017}]%
        {fib-oopsla-2017}
\bibfield{author}{\bibinfo{person}{Benjamin~P. Wood}, \bibinfo{person}{Man
  Cao}, \bibinfo{person}{Michael~D. Bond}, {and} \bibinfo{person}{Dan
  Grossman}.} \bibinfo{year}{2017}\natexlab{}.
\newblock \showarticletitle{{Instrumentation Bias for Dynamic Data Race
  Detection}}.
\newblock \bibinfo{journal}{\emph{PACMPL}} \bibinfo{volume}{1},
  \bibinfo{number}{OOPSLA}, Article \bibinfo{articleno}{69}
  (\bibinfo{date}{Oct.} \bibinfo{year}{2017}), \bibinfo{numpages}{31}~pages.
\newblock
\showISSN{2475-1421}


\bibitem[\protect\citeauthoryear{Wood, Ceze, and Grossman}{Wood
  et~al\mbox{.}}{2014}]%
        {lard}
\bibfield{author}{\bibinfo{person}{Benjamin~P. Wood}, \bibinfo{person}{Luis
  Ceze}, {and} \bibinfo{person}{Dan Grossman}.}
  \bibinfo{year}{2014}\natexlab{}.
\newblock \showarticletitle{{Low-Level Detection of Language-Level Data Races
  with LARD}}. In \bibinfo{booktitle}{\emph{ASPLOS}}.
  \bibinfo{pages}{671--686}.
\newblock
\showISBNx{978-1-4503-2305-5}


\bibitem[\protect\citeauthoryear{Wu, Ouyang, Zhou, Zhang, Liu, and Zhang}{Wu
  et~al\mbox{.}}{2020}]%
        {simulee-icse-2020}
\bibfield{author}{\bibinfo{person}{Mingyuan Wu}, \bibinfo{person}{Yicheng
  Ouyang}, \bibinfo{person}{Husheng Zhou}, \bibinfo{person}{Lingming Zhang},
  \bibinfo{person}{Cong Liu}, {and} \bibinfo{person}{Yuqun Zhang}.}
  \bibinfo{year}{2020}\natexlab{}.
\newblock \showarticletitle{{Simulee: Detecting CUDA Synchronization Bugs via
  Memory-Access Modeling}}. In \bibinfo{booktitle}{\emph{ICSE}}.
  \bibinfo{pages}{937--948}.
\newblock
\showISBNx{9781450371216}


\bibitem[\protect\citeauthoryear{Yu, Rodeheffer, and Chen}{Yu
  et~al\mbox{.}}{2005}]%
        {racetrack}
\bibfield{author}{\bibinfo{person}{Yuan Yu}, \bibinfo{person}{Tom Rodeheffer},
  {and} \bibinfo{person}{Wei Chen}.} \bibinfo{year}{2005}\natexlab{}.
\newblock \showarticletitle{{RaceTrack: Efficient Detection of Data Race
  Conditions via Adaptive Tracking}}. In \bibinfo{booktitle}{\emph{SOSP}}.
  \bibinfo{pages}{221--234}.
\newblock


\bibitem[\protect\citeauthoryear{Zhang, Jung, and Lee}{Zhang
  et~al\mbox{.}}{2017}]%
        {prorace}
\bibfield{author}{\bibinfo{person}{Tong Zhang}, \bibinfo{person}{Changhee
  Jung}, {and} \bibinfo{person}{Dongyoon Lee}.}
  \bibinfo{year}{2017}\natexlab{}.
\newblock \showarticletitle{{ProRace: Practical Data Race Detection for
  Production Use}}. In \bibinfo{booktitle}{\emph{ASPLOS}}.
  \bibinfo{pages}{149--162}.
\newblock
\showISBNx{978-1-4503-4465-4}


\bibitem[\protect\citeauthoryear{Zhang, Lee, and Jung}{Zhang
  et~al\mbox{.}}{2016}]%
        {txrace}
\bibfield{author}{\bibinfo{person}{Tong Zhang}, \bibinfo{person}{Dongyoon Lee},
  {and} \bibinfo{person}{Changhee Jung}.} \bibinfo{year}{2016}\natexlab{}.
\newblock \showarticletitle{{TxRace: Efficient Data Race Detection Using
  Commodity Hardware Transactional Memory}}. In
  \bibinfo{booktitle}{\emph{ASPLOS}}. \bibinfo{pages}{159--173}.
\newblock
\showISBNx{978-1-4503-4091-5}


\bibitem[\protect\citeauthoryear{Zheng, Ravi, Qin, and Agrawal}{Zheng
  et~al\mbox{.}}{2011}]%
        {grace-ppopp-2011}
\bibfield{author}{\bibinfo{person}{Mai Zheng}, \bibinfo{person}{Vignesh~T.
  Ravi}, \bibinfo{person}{Feng Qin}, {and} \bibinfo{person}{Gagan Agrawal}.}
  \bibinfo{year}{2011}\natexlab{}.
\newblock \showarticletitle{{GRace: A Low-Overhead Mechanism for Detecting Data
  Races in GPU Programs}}. In \bibinfo{booktitle}{\emph{PPoPP}}.
  \bibinfo{pages}{135--146}.
\newblock
\showISBNx{9781450301190}


\bibitem[\protect\citeauthoryear{Zheng, Ravi, Qin, and Agrawal}{Zheng
  et~al\mbox{.}}{2014}]%
        {gmrace-tpds-2014}
\bibfield{author}{\bibinfo{person}{Mai Zheng}, \bibinfo{person}{Vignesh~T.
  Ravi}, \bibinfo{person}{Feng Qin}, {and} \bibinfo{person}{Gagan Agrawal}.}
  \bibinfo{year}{2014}\natexlab{}.
\newblock \showarticletitle{{GMRace: Detecting Data Races in GPU Programs via a
  Low-Overhead Scheme}}.
\newblock \bibinfo{journal}{\emph{{IEEE} TPDS}} \bibinfo{volume}{25},
  \bibinfo{number}{1} (\bibinfo{date}{Jan.} \bibinfo{year}{2014}),
  \bibinfo{pages}{104--115}.
\newblock
\showISSN{1045-9219}


\bibitem[\protect\citeauthoryear{Zhou, Teodorescu, and Zhou}{Zhou
  et~al\mbox{.}}{2007}]%
        {zhou-hard}
\bibfield{author}{\bibinfo{person}{Pin Zhou}, \bibinfo{person}{Radu
  Teodorescu}, {and} \bibinfo{person}{Yuanyuan Zhou}.}
  \bibinfo{year}{2007}\natexlab{}.
\newblock \showarticletitle{{HARD: Hardware-Assisted Lockset-based Race
  Detection}}. In \bibinfo{booktitle}{\emph{HPCA}}. \bibinfo{pages}{121--132}.
\newblock
\showISBNx{1-4244-0804-0}


\end{thebibliography}

\iftoggle{acmFormat}{
    \pagebreak
}{}

\end{document}